\documentclass[floatfix,pre,preprint,floats,aps]{revtex4}
\usepackage{graphicx}
\begin{document}

\title{Exact Solution of a Field Theory Model of
Frontal Photopolymerization\footnote{Contribution of the National
Institute of Standards and Technology, not subject to US
copyright.}}

\author{James A. Warren,\footnote{Correspondence:
\href{mailto:tojames.warren@nist.gov}{james.warren@nist.gov},
\href{mailto:joao.cabral@nist.gov}{joao.cabral@nist.gov} and
\href{mailto:jack.douglas@nist.gov}{jack.douglas@nist.gov}}
Jo\~{a}o T. Cabral,$^\ddag$ and Jack F. Douglas$^\ddag$}

\address{Metallurgy$^\dag$ and Polymers$^\ddag$ Divisions, National
  Institute of Standards and Technology,
Gaithersburg MD 20899}

\begin{abstract}
Frontal photopolymerization (FPP) provides a versatile method for
the rapid fabrication of solid polymer network materials by
exposing photosensitive molecules to light. Dimensional control of
structures created by this process is crucial in applications
ranging from microfluidics and coatings to dentistry, and the
availability of a predictive mathematical model of FPP is needed
to achieve this control.  Previous work has relied on numerical
solutions of the governing kinetic equations in validating the
model against experiments because of the intractability of the
governing non-linear equations. The present paper provides
\textit{exact} solutions to these equations in the general case in
which the optical attenuation decreases (photobleaching) or
increases (photodarkening) with photopolymerization. These exact
solutions are of mathematical and physical interest because they
support traveling waves of polymerization that propagate
logarithmically or linearly in time, depending on the evolution of
optical attenuation of the photopolymerized material.
\end{abstract}

\maketitle


\section{Introduction}
\label{intro}

Photopolymerization is a common method of rapidly forming solid
network polymer materials and it is possible to create intricate
three-dimensional structures by selectively polymerizing
photosensitive materials through masks opaque to light. The
conversion process from a liquid to a solid does not occur
uniformly in this fabrication technique because of the attenuation
of light within the photopolymerizable material (PM) and this
process is normally accompanied by non-uniform monomer-to-polymer
conversion profiles perpendicular to the illuminated surface
\cite{odian,Fouassier,fourab,decker1,decker2}. Physically, these
conversion profiles propagate as \textit{traveling waves} of
network solidification that invade the unpolymerized medium
exposed to radiation (generally ultraviolet light, UV) if the
process occurs in the presence of strong optical attenuation and
limited mass and heat transfer. The frontal aspect of the
polymerization process is apparent in the photopolymerization of
thick material sections and has counterparts in degradation
(including discoloration) processes in polymer films exposed to UV
radiation, where the breaking of chemical bonds  rather than their
formationis often the prevalent physical process.

Frontal photopolymerization (FPP) is utilized in diverse
fabrication processes, ranging from photolithography of
microcircuits to dental restorative and other biomedical
materials, and numerous coatings applications (paints and
varnishes, adhesives and printing inks) \cite{decker1,decker2}. We
have recently explored the use of FPP in the fabrication of
microfluidic devices \cite{harrison, cabral2004, tao, cygan,
hudson}.

We emphasize that FPP is a distinct mode of polymerization from
thermal (TFP) and isothermal (IFP) frontal polymerization, which
involve \textit{autocatalytic reactions}. While these
polymerization methods also involve wavelike polymerization
fronts, the front propagation is sustained by the thermal energy
released from an exothermic polymerization reaction. This
self-propagating frontal growth can be initiated by a localized
heat source (TFP) of by a polymer network seed (IFP) and has been
reviewed by Pojman \emph{et al.} \cite{pojman1,pojman2,pojman3}.

Given the complexity of the chemical reactions involved in FPP, a
`minimal' field theoretic model of this process was introduced in
previous work based on physical observables relevant to the
fabrication process \cite{cabral2004, cabral2005}. Specifically,
this FPP model concerns itself with two basic front properties and
their evolution in space and time: (1) the position of the
solid/liquid front, which defines the patterned height and (2) the
light transmission of the PM layer. This formulation naturally
leads to a system of coupled partial differential equations
involving two coupled field variables, the extent of
monomer-to-polymer conversion $\phi(x,t)$ and the light
attenuation $Tr(x,t)$ as a function of the distance from the
illuminated surface $x$ and time $t$.

Before describing our mathematical model, we briefly illustrate
the physical nature of FPP through experiments on a model UV
polymerizable material, described in Section \ref{expt} and
discussed in Section \ref{frontal}. The derivation of this model
is reviewed in Section \ref{model} and Section \ref{exact}
presents exact solutions of these non-linear equations.

\section{Experimental}
\label{expt}

\begin{figure}[htbpH]
\begin{center}
\leavevmode
\includegraphics[width=13cm]{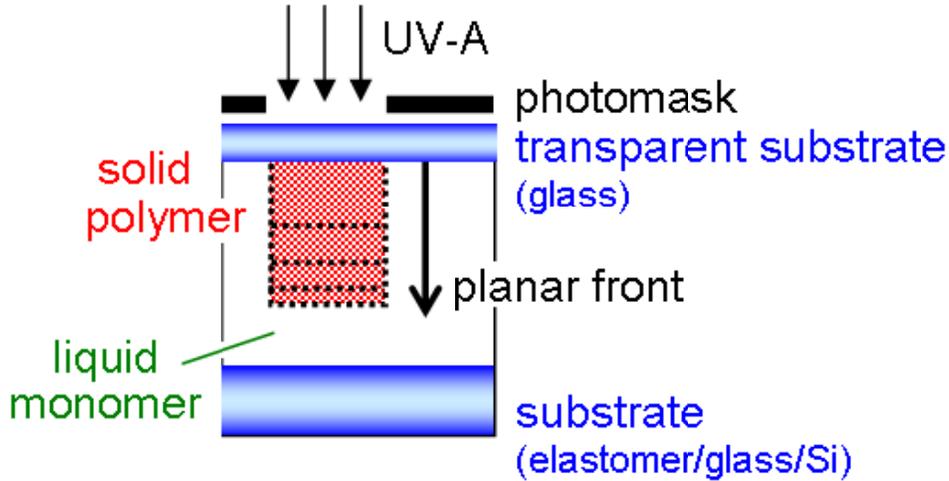}
\end{center}
\caption{Schematic of the frontal photopolymerization (FPP),
depicting a monomer-to-polymer conversion front induced by light
exposure moving towards the bulk polymerizable material (PM). Our
experimental setup consists of a collimated UV source (365 nm), a
photomask, and a PM confined between two surfaces, typically glass
and an elastomer sheet. } \label{fig-schem}
\end{figure}

The photopolymerization experimental setup \cite{disclaimer}
consists of a collimated light source, a photomask, a polymer
photoresist and a substrate, as depicted in Fig. \ref{fig-schem}.
We choose a multifunctional thiol-ene formulation (NOA81, Norland
Products, NJ) as the photopolymerizable material (PM) for this
study. This optically clear, liquid PM functions as a negative
photoresist and cures under 365 nm ultraviolet light (UVA) into a
hard solid (Shore D durometer 90 and $\sim$ 1 GPa modulus).
Moreover, thiol-enes polymerize rapidly at ambient conditions
(with minimal oxygen inhibition) and achieve large depths of cure
\cite{jacobine,Cramer02,Cramer03a,reddy}. In previous work, we
have characterized the kinetics of FPP of these systems as a
function of PM composition, temperature and nanoparticle loading
\cite{cabral2004, cabral2005}.

The liquid PM was poured into an elastomeric
(polydimethylsiloxane, Sylgard 184, Dow Corning) gasket and
covered with a plasma-cleaned glass slide (Corning 2947). The
oxygen plasma was an Anatech-SP100 operating at 80 Pa (600 mTorr), with 60
W for 3 min. Photomasks were printed on regular acetate sheet
transparencies (CG3300, 3M) using a 1200 dots per inch HP Laserjet 8000N
printer. The mask consisted of a square array of large posts (2 mm
$\times$ 2 mm) and was placed directly over the top glass slide.
An aluminum shutter was placed over the specimen and moved
manually, controlling the exposure time of each post. The light
source was a Spectroline SB-100P flood lamp, equipped with a 100
Watt Mercury lamp (Spectronics), placed at a variable distance
(100's of mm) from the specimen to adjust the incident intensity.
The light intensity was measured with a Spectroline DIX-365A UV-A
sensor and DRC-100X radiometer (both Spectronics) with 0.1
\(\mu\)W/mm$^2$ (10
\(\mu\)W/cm$^2$) resolution. The UV dose administered to each
patterned post was calculated as the product of the incident light
intensity $I_0\equiv I(x=0)$, light transmission $Tr$ of
the mask ($\sim$ 80 \%) and glass slide ($\sim$ 94 \%), and
exposure time $t$, as $UV dose \equiv Tr I_0 t$ ; $x$ is
depth distance normal to the surface in the PM.
Photopolymerization was carried out under a fume hood at 30
$^\circ$C, with incident light intensity of (2 and 10)
\(\mu\)W/mm$^2$; a wide UV dose window covering 0.04 mJ/mm$^2$ to 180
mJ/mm$^2$ was investigated.

Upon UV light exposure, imaged areas become insoluble to selective
solvents ethanol and acetone, which are used to develop the
pattern. Compressed air and a succession of alternating
ethanol/acetone rinses are employed until the unpolymerized
material is thoroughly removed. The resulting pattern has well
defined dimensions but is still a `soft' solid. A flood UV
exposure (for about 50 times the patterning dose), completes the
crosslinking process of the material into a hard solid, largely
preserving its dimensions.  The topography of the resulting
photopolymerized structure was mapped by stylus profilometry,
using a Dektak 8 profilometer (Veeco, CA), equipped with a 12.5
\(\mu\)m stylus and operating at 10 mg force. For post heights
beyond the profilometer 1 mm limit, a caliper (Digit-cal MK IV,
Brown \& Sharpe) was utilized. Measurement uncertainty ranged from
5 \% to 10 \%, depending on the pattern height. A typical
profilometer scan of two arrays of posts exposed to increasing UV
doses is shown in Fig. 2. The resulting patterned dimensions range
from approximately 70 to 1000 \(\mu\)m in height.

\begin{figure}[htbpH]
\begin{center}
\leavevmode
\includegraphics[width=13cm]{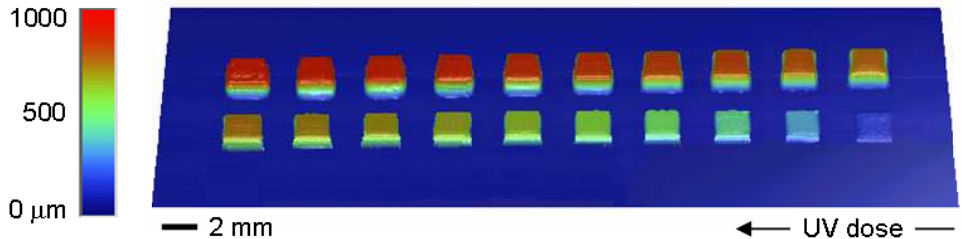}
\end{center}
\caption{Topography map of an array of FPP squares obtained by
stylus profilometry. The exposure time for each square was defined
by a shutter system and was varied \textit{linearly} in 30 s
intervals, totaling 10 min. The resulting heights $h(t)$, however,
increase in a strongly non-linear fashion, apparently leveling off
at long exposures. The incident intensity was 1.8 \(\mu\)W/mm$^2$
and the UV dose window sampled was (0.05 to 10) mJ/mm$^2$. }
\label{fig-profi}
\end{figure}

In order to explore the spatio-temporal variation of the light
intensity upon photocuring, a second series of experiments were
devised. The transmission of PM samples of different thickness was
monitored as a function of time during the conversion process. The
PM was confined between transparent glass slides with spacers of
defined thickness; this assembly was placed between the UV source
and the radiometer and the transmitted light intensity was
recorded as a function of time. Sample thickness was limited to 1
mm due to light attenuation and sensor sensitivity to the actinic
wavelength. The effective sample transmission $Tr(x,t)$ was
obtained from the recorded intensity $I(x,t)$ and the Beer-Lambert
relation $Tr(x,t) \equiv (I(x,t)/I_0)/ Tr(glass)^{2} = \exp \left[
-\bar{\mu}(x,t)x \right]$, after subtracting the attenuation due
to the glass slides (2 $\times$ 1 mm); $x$  is the sample
thickness (a constant in this experiment) and $t$  is the
exposure time.

\section{Frontal Polymerization Induced by Light}
\label{frontal}

We first establish the basic nature of the frontal
photopolymerization (FPP) based on experimental evidence. The
propagation of a planar monomer-to-polymer conversion front,
emanating from the illuminated surface, is depicted in Fig.
\ref{fig-schem}. A topographic map of arrays of FPP fronts
measured by profilometry is shown in Fig. \ref{fig-profi}. The
interface between the polymerized solid and the liquid
pre-polymer, characteristic of frontal polymerization, is evident
after `development' (selective washing away of the unpolymerized
material) of the pattern. The height dependence of exposure dose
(the product of exposure time $t$ and light intensity $I_0$) was
obtained from a series of experiments and characterizes the FPP
frontal kinetics. Results for the PM studied, for a light dose
window of a few millijoues per square centimeter to 20 J/cm$^2$, at 30
$^\circ$C are shown in Fig. 3a. We define `front position' $h(t)$  in a
straightforward way as the measured thickness of the solidified
material after UV exposure and development (washing away the
unsolidified PM). This criterion is a natural choice for rapid
prototyping and fabrication using FPP. Also, in practical
applications, it is useful to express results in terms of light
dose, rather than exposure time. The validity of interchanging
dose and $t$ depends on the reaction kinetics independence of
$I_0$, which applies to the PM in the conditions studied
\cite{cabral2004}.

\begin{figure}[htbpH]
\begin{center}
\leavevmode
\includegraphics[width=11cm]{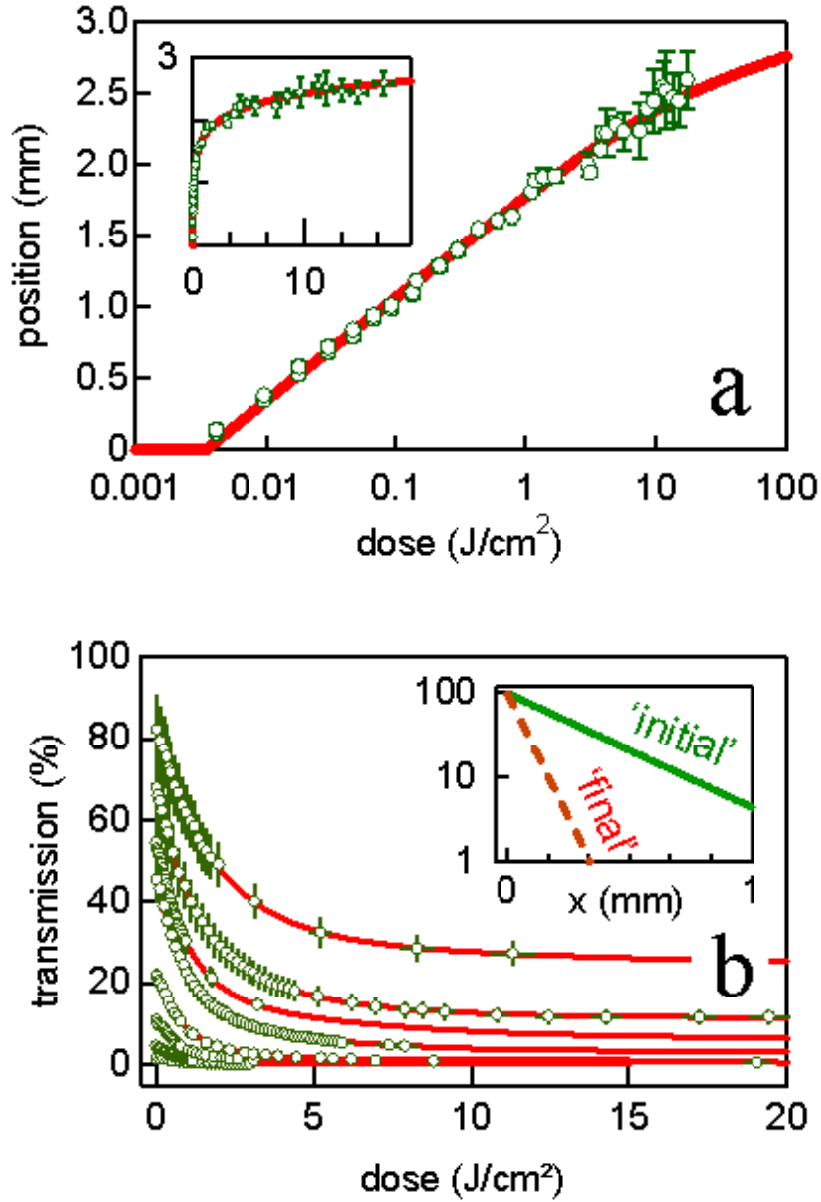}
\end{center}
\caption{Experimental FPP results for an illustrative `partial
photo-darkening' polymerization. (a) Front position dependence on
UV dose (light intensity × exposure time) showing an initial
logarithmic dependence followed by a crossover. The inset is a
linear plot. (b) Optical transmission (up to 365 nm) variation during
photocuring for PM samples of constant thickness. The inset
depicts the log transmission as a function of thickness for the
`initial' (before conversion) and final (`full conversion') stages
of photopolymerization, where the simple Beer-Lambert law holds,
yielding the asymptotic $\mu_0$ and $\mu_\infty$ attenuation
coefficients.  } \label{fig-exp}
\end{figure}

The optical transmission of this specific PM decreases during
photocuring and this process is captured in Fig. 3b for a series
of specimens with different thickness. There is clearly a drop in
$Tr$ upon photopolymerization indicating partial photodarkening.
The figure inset shows the thickness-dependent transmission before
(`initial') and after (`final') a long UV exposure (until
\textit{Tr} reaches a plateau), in the usual Beer-Lambert
representation. Other photoresists `photobleach' during the
process, due to consumption of a strongly absorbing species
(generally the photoinitiator), or may remain virtually
`invariant' (with constant light transmission) upon conversion.
The experimental results presented in Fig. 3 characterize the
general nature of FPP and illustrate the kinetics of its
observables, front position $h(t)$ and transmission $Tr(x,t)$, in
a `photodarkening' material.

\section{Frontal Photopolymerization (FPP) Model}
\label{model}

Photopolymerization begins with the absorption of light, which
generates the reactive species responsible for chain initiation.
The addition of a strongly light-absorbing photoinitiator modifies
the optical properties of the medium and its consumption in the
course of network formation, in conjunction with network formation
and the formation of photopolymerization by-products, leads to an
evolving optical attenuation. The consumption of the
photoinitiator alone can be expected to lead to a reduction of the
optical attenuation in the UV frequency range (`photobleaching'),
but the resulting polymer network can have an increased optical
attenuation so that the net optical attenuation can
\emph{increase} upon photopolymerization (`photodarkening').
Moreover, the addition of nanoparticle additives will also change
the optical properties of the medium from those of the unfilled
material in a non-trivial fashion.\cite{cabral2005} We thus develop a model of
photopolymerization that does not presume either photobleaching or
photodarkening as a general consequence of photopolymerization.
The nature of the polymerization front development has distinct
features in these physical situations that we discuss in separate
sections below after summarizing our general model.

The kinetic model of FPP \cite{cabral2004, cabral2005} conceives
of the photopolymerization process in terms of a coarse-grained
field theoretic perspective. The state of the material is assumed
to be characterized by field variables that describe the extent
to which the material is polymerized and the spatially and
temporally dependent optical attenuation evolves in response to
the photopolymerization process. While this model has mathematical
similarities with classic theories of photo-polymerization
\cite{Wegscheider,Mauser}, it directly focuses on observable
properties of FPP rather than the concentration of the various
chemical species involved.  The main variables of interest in the
kinetic model are the FPP front position $h(t)$, as defined, for
example, by the solid/liquid interface, the light transmission
$Tr(x,t)$ of the PM layer and the optical attenuation constants
($\mu_0$, $\mu_\infty$) of the monomer and the fully converted
material, respectively. The \textit{extent of polymerization}
$\phi (x,t)$ is then introduced as an `order parameter' describing
the extent of conversion of the growing polymerization front. The
field variable $\phi (x,t)$ describes the average ratio of
photopolymerized to unpolymerized material at a depth $x$ (the
illuminated surface defines the coordinate origin) into the PM and
satisfies the limiting relations $\phi(x,t\rightarrow 0)=0$ (no
polymer) and $\phi(x,t \rightarrow \infty) =1$ (full
polymerization) for all $ x
> 0$. The second field variable $Tr(x,t)$ describes the optical
transmission of the photopolymerizable medium of thickness $x$ at
time $t$. This coarse-grained description of the
photopolymerization front propagation has analogies with
phase-field descriptions of ordering processes such a
crystallization and dewetting where propagating fronts are also
observed \cite{warren95,ferreiro02}.

The evolution of the photopolymerization process is modeled by
introducing appropriate rate laws for the specified minimal set of
field variables \cite{cabral2004, cabral2005}. The rate of change
of $\phi (x,t)$ is taken to be proportional to the optical
transmission $Tr(x,t)$, the amount of material available for
conversion  and the reaction conversion rate $K$,
\begin{equation}
\frac{\partial\phi(x,t)}{\partial t}= K \left[ 1- \phi(x,t)
\right] Tr(x,t) \label{model-phi}
\end{equation}

Once photopolymerization has commenced, the material is considered
to be a two-component system (consisting of reacted and unreacted
material) whose components do not generally have the same optical
attenuation coefficient $\mu$. The
\textit{non-uniformity} of the conversion profile will generally
give rise to an effective attenuation factor $\bar{\mu}(x,t)$,
which depends on thickness during conversion. Only before
photocuring and near full conversion $\bar{\mu}(x,t)$ becomes
constant. In our mean-field model, we postulate that the material can
be described using a  spatially varying and temporally evolving average
optical attenuation,
\[ \bar{\mu} (x,t)\equiv \mu_0 \left[ 1-
\phi(x,t) \right] + \mu_{\infty} \phi(x,t), \] where ($\mu_0$) and
($\mu_\infty$) are the attenuation coefficients of the unexposed
monomer  and fully polymerized material, respectively. The
variation leads to an evolution in the light intensity (or
transmission) profile with depth according to the generalized
Beer-Lambert relation,
\begin{equation}
\frac{\partial Tr(x,t)}{\partial x}= - \bar{\mu} (x,t) Tr(x,t),
\label{model-int}
\end{equation}
where the usual Beer-Lambert law for a homogeneous material,
$Tr(x,t)=\exp ( -\bar{\mu} x)$, is recovered for short and long
times as $\bar{\mu}(x,t \rightarrow 0) = \mu_0$ and $\bar{\mu}(x,t
\rightarrow \infty) = \mu_\infty$.

Specific boundary conditions must be specified in order to solve
such differential equations. Initially $\phi(x,0)=0$, while at the
incident surface of the sample ($x=0$), we have no attenuation,
thus $Tr(0,t)=1$. These are sufficient to determine unique
solutions to Eqs. (\ref{model-phi}) and (\ref{model-int}). We
should also note that we can quickly solve Eq. (\ref{model-phi})
when $x=0$ to obtain
\begin{equation}
  \label{eq:phi0}
  \phi_0(t)\equiv\phi(0,t)=1-\exp(-Kt),
\end{equation}
an expression for the polymerized fraction at the edge of the sample
that is {\sl independent} of all model parameters except $K$.

The idealization of  FPP  evolution modeled by Eqs.
(\ref{model-phi}) and (\ref{model-int}) neglects the fact that
numerous chemical components are actually generated in the course
of photopolymerization and ignores the presence of additives and
impurities that are often present in the photopolymerizable
material. Additionally, it assumes simple chemical kinetics,
defined by a single \emph{constant} $K$. Thus, it is not clear a
priori whether such a simple order parameter treatment of FPP is
suitable. Judgement of the adequacy of our approach must be
decided by comparison to measurements performed over a wide range
of conditions. We next consider the final basic observable
property of the FPP process, the position of the
photopolymerization front.

As in ordinary gelation, we can expect solidification to occur
once $\phi$ exceeds a certain `critical conversion fraction'
$\phi_c$ ($\ll$1). Since the liquid material can be simply washed
away after any exposure time, the height $h(t)$ at which
$\phi(x,t)= \phi_c$ indicates the surface of the photopolymerized
material after curing and washing. This defines the position of
FPP front in a concrete way and we adopt it below. Our previous
measurements have shown that $\phi_c$ tends to be rather small
[$\phi_c \sim \emph{O}(0.01)$] in our thiol-ene photopolymerizable
material \cite{cabral2004,cabral2005} and this property is
expected to be rather general. A small $\phi_c$ can be understood
from the fact that solidification in polymerizing materials
\cite{MCKENNA} (e.g., `superglue') normally involves a combination
of glass formation and gelation, since the glass transition
temperature strongly increases upon polymerization of a low
molecular weight monomer. Accordingly, we adopt the representative
value $\phi_c=0.02$ in our discussion below.

Equations (\ref{model-phi}) and (\ref{model-int}) define a system
of non-linear partial differential equations whose solution
depends on three material parameters: the short and long-time
attenuation coefficients, as well as the conversion rate $K$. The
former two parameters can be measured independently with a series
of transmission measurements of unpolymerized and fully
polymerized specimens of different thicknesses. $K$ is determined
by the polymerization chemistry and is a structural variable, yet
both can be obtained as fitting parameters. The former has been
the focus of much of the previous research \cite{rytov, terrones1,
terrones2, terrones3, terrones4, Ivanov, goodner02, obrien,
Miller, belk, Cramer02, Cramer03a, reddy, Wegscheider, Mauser},
and is not discussed in the present paper.

The coupled \textit{non-linear} differential Eqs.
(\ref{model-phi}) and (\ref{model-int}) have not yet been solved
analytically, apart from special limits that are briefly
summarized in the next section.  These exactly solvable cases
include `total photobleaching' where $\mu_0 > 0$ and $\mu_\infty =
0$ and `photo-invariant polymerization' in which the optical
properties of the medium do not change in the course of
polymerization (i.e., $\mu_0 = \mu_\infty \equiv \bar{\mu}$).
Front propagation is quite different in these different physical
situations and we briefly describe the nature of FPP in these
limiting cases, and then explore the full solution in some other
physically relevant cases, where we  identify those basic features
of FPP that can be recognized experimentally. Rytov \emph{et al.}
\cite{rytov}\ is one of few previous papers to study these
different types of FPP, both by analytic modelling and experiment.
This work, however, had to introduce rough approximations to
obtain estimates of front properties.

\section{Exact Formal Solution of Kinetic Equations in Limiting Cases}
\label{exact}

\subsection{Total Photobleaching \\
($\mu_0 > 0$ and $\mu_\infty = 0$)}

The initiator of the photopolymerization reaction often absorbs
light strongly and the absorption of radiation can expected to
lead to a reduction of the optical attenuation upon UV radiation
through the chemical degradation of this reactive species. If this
was the only species contributing to the optical attenuation of
the medium, then the photopolymerized material would become
increasingly transparent to light, becoming perfectly transparent
to the radiation at infinite times. This is evidently an idealized
model of photopolymerized materials, but most theoretical
discussions of photopolymerization \cite{decker1, decker2, rytov,
terrones1, terrones2, terrones3, terrones4, Ivanov, goodner02,
obrien, Miller, belk}\ are restricted to this limiting case based
on the assumption that the PM initiator dominates the optical
attenuation.

The case of perfect optical absorption is one of the few cases in
which an exact solution can be expressed in terms of elementary
functions, and this solution is instructive into basic features of
FPP. In this case, the PM has a positive attenuation constant
($\mu_0>$ 0) and the attenuation of the polymerized material
equals, $\mu_\infty = 0$. In this case, Eqs. (\ref{model-phi}) and
(\ref{model-int}) can be easily solved to find that the conversion
fraction $\phi (x,t)$ for perfect photobleaching equals \cite{cabral2005}
\begin{equation}
\phi(x,t) =\frac{1-\exp(-Kt)}{1-\exp(-Kt)+\exp(\mu_0x-Kt)}.
\label{eq-perfect}
\end{equation}
Note that this expression reduces to Eq. (\ref{eq:phi0}) when
$x=0$, and that the conversion fraction is defined solely  for
$x>0$. Eq. (\ref{eq-perfect}) was obtained long ago by Wegscheider
\cite{Wegscheider}, but the physical interpretation of these
equations differs in his treatment which models the concentration
of reactive species, rather than the extent of
photopolymerization.

Equation (\ref{eq-perfect}) can be written equivalently in terms
of the coordinate $z$  moving with the front as,
\begin{eqnarray}
\phi(z,t)  =  1/ \left[ 1+\exp(\mu_0 z)\right]=
\frac{1}{2}\left(1+\tanh\left(\frac{\mu_0 z}{2}\right)\right)
\\
z = x-x_f,\ x_f  =  \left[ Kt + \ln \left[ 1 - \exp(-Kt)
\right]\right] / \mu_0, \label{eq-perfectm}
\end{eqnarray}
where $x_f$ is the inflection point of the front that propagates
in space as the front advances. This position can also be
identified in this model by a mathematically equivalent condition
$\phi=1/2$, and the front position can thus can be defined by a
(unique) maximum in $-\partial\phi(x,t)/\partial x=\phi_x$,
\begin{equation}
  \label{eq:defxp}
  \left.\frac{\partial^2\phi}{\partial x^2}\right|_{x_f}=0.
\end{equation}

The position $x_f$ is particularly applicable as a definition of
the interface location if optical methods are used to probe the
position of the front. Alternatively, as described in the previous
section, it is sometimes more useful to define the front position
by a `critical' value of the order parameter $\phi(x,t)=\phi_c$
(e.g., value of $\phi$ at which the material becomes a solid).
This front definition \cite{cabral2004, cabral2005, Hirose} leads
to a travelling wave solution whose displacement also obeys
Eq. (\ref{eq-perfectm}).

Indeed, if we define a new coordinate $z_h=x-h(t)$, and insist
that $\phi(z_h=0)=\phi_c$, we determine $h(t)$ as,
\begin{equation}
  \label{eq:hsc1}
  h(t)=x_f+\frac{1}{\mu_0}\ln\left(\frac{1}{\phi_c}-1\right).
\end{equation}

Using the representative value of $\phi_c=0.02$ introduced above,
we plot $h(t)$ in Fig. \ref{fig:xpsc1and2}. The offset between our
two interface position choices is then $\mu_0(h-x_f)\approx3.892$,
for this example.

\begin{figure}[htbpH]
   \begin{center}
      \includegraphics[width=13cm]{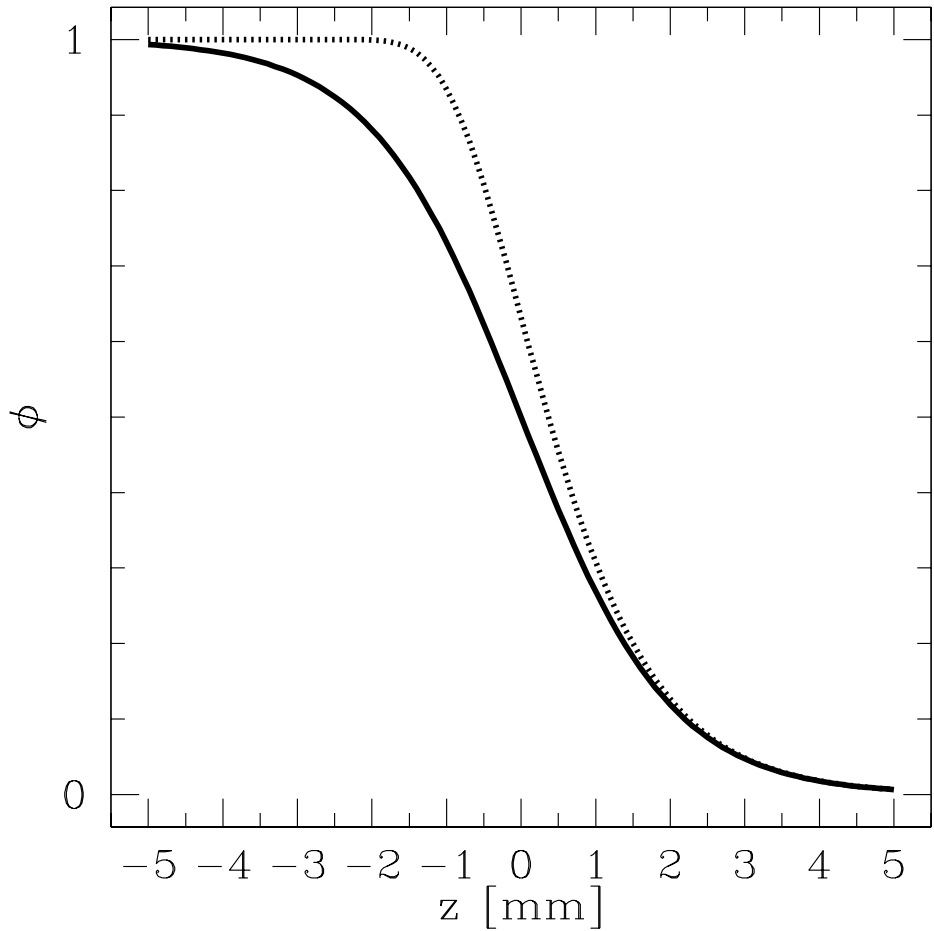}
      \caption{Conversion fraction $\phi$ as a function of z for both total
        photobleaching (solid) and photo-invariant polymerization
        (dotted), with $\mu_0=1.0$ mm$^{-1}$.}
   \label{fig:phisc1and2}
   \end{center}
\end{figure}

\begin{figure}[htbpH]
   \begin{center}
      \includegraphics[width=13cm]{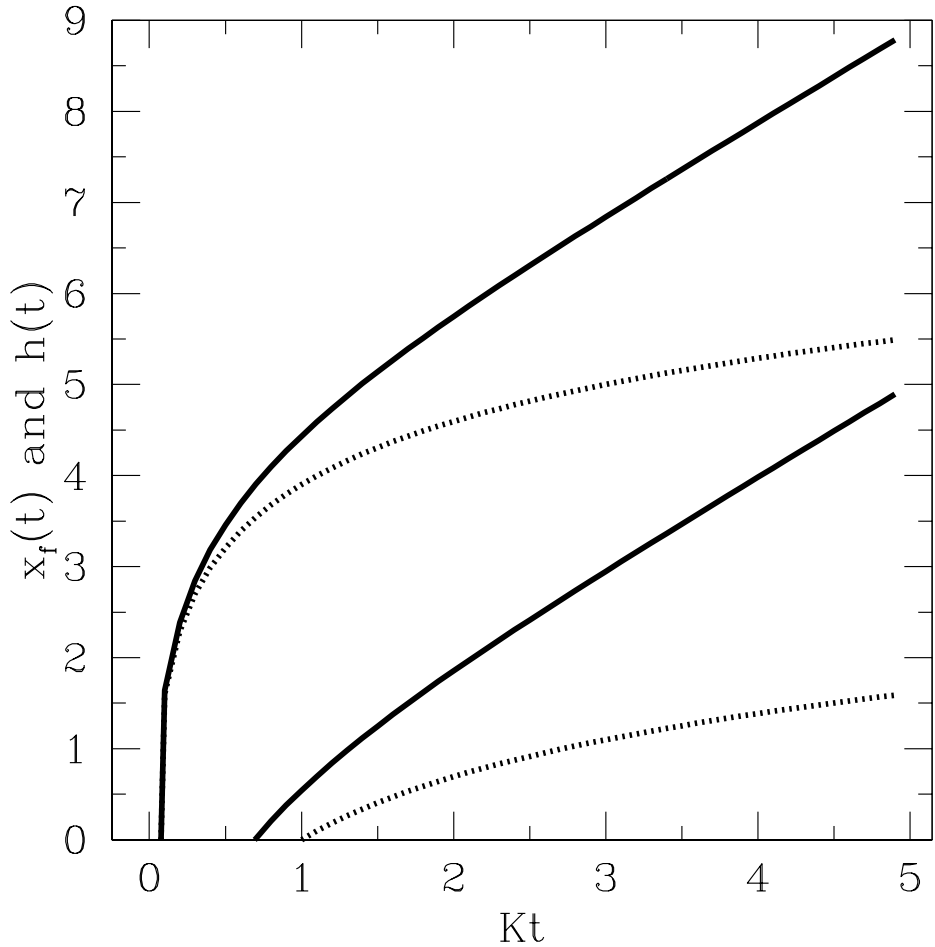}
      \caption{Plots of $x_f$ [mm] (emerging after an induction time $Kt\approx1$)
        and $h(t)$ [mm] (emerging with little induction time at $Kt\rightarrow0$) as a
        function of $Kt$ for both total
        photobleaching (solid) and photo-invariant polymerization
        (dotted), for $\mu_0=1.0$ mm$^{-1}$. At late times the total
        photobleaching position has a linear slope, corresponding to
        a front velocity of $K/\mu_0$}
   \label{fig:xpsc1and2}
   \end{center}
\end{figure}

Equation (\ref{eq-perfectm}) implies that $\phi(x,t)$ evolves as a
\textit{propagating sigmoidally-shaped front} whose position is
defined by $x_f$. Since this profile will be compared with $\phi$
profiles for the general solution of Eqs. (1) and (2) below, we
plot $\phi(z)$ in Figure \ref{fig:phisc1and2} (the photo-invariant
profile is discussed in the following section).

The position of this front $x_f$ (defined here by the inflection
point, or $\phi=1/2$) is shown in Fig. \ref{fig:xpsc1and2}. At
long times ($t\gg K^{-1}$), the front translates linearly in time
with a constant velocity $K/\mu_0$. Linear front propagation has
commonly been reported in experimental studies of FPP kinetics
(e.g., \cite{rytov}).

At early times the position of the inflection point lies outside
the polymerizing sample. Specifically, Eq. (\ref{eq:phi0}) implies
$\phi(0,t)=1-\exp(-Kt)$, which can be less than $\phi=1/2$, the
value of $\phi$ at the inflection point. The inflection point
appears after an \emph{induction time},
\begin{equation}
\tau =\frac{\ln2}{K}, \label{eq-perfectt}
\end{equation}
which explains the intercept of the interface position shown in
Fig. \ref{fig:xpsc1and2}.

From our definition of the position of the FPP front,
the width of the front $\xi$ can be correspondingly defined as
the reciprocal of the magnitude of $\phi_x$ at the front position,
\begin{equation}
\xi \equiv 1/ \left| \phi_x(x_f) \right| \label{eq-xi}
\end{equation}
This definition is suitable for any symmetric front shape for
which $\phi (x,t) \approx 1/2 $ at the inflection point and we
note that $\phi (x_f,t)$ exactly equals $1/2$ for total
photobleaching.

\begin{figure}[htbpH]
   \begin{center}
      \includegraphics[width=13cm]{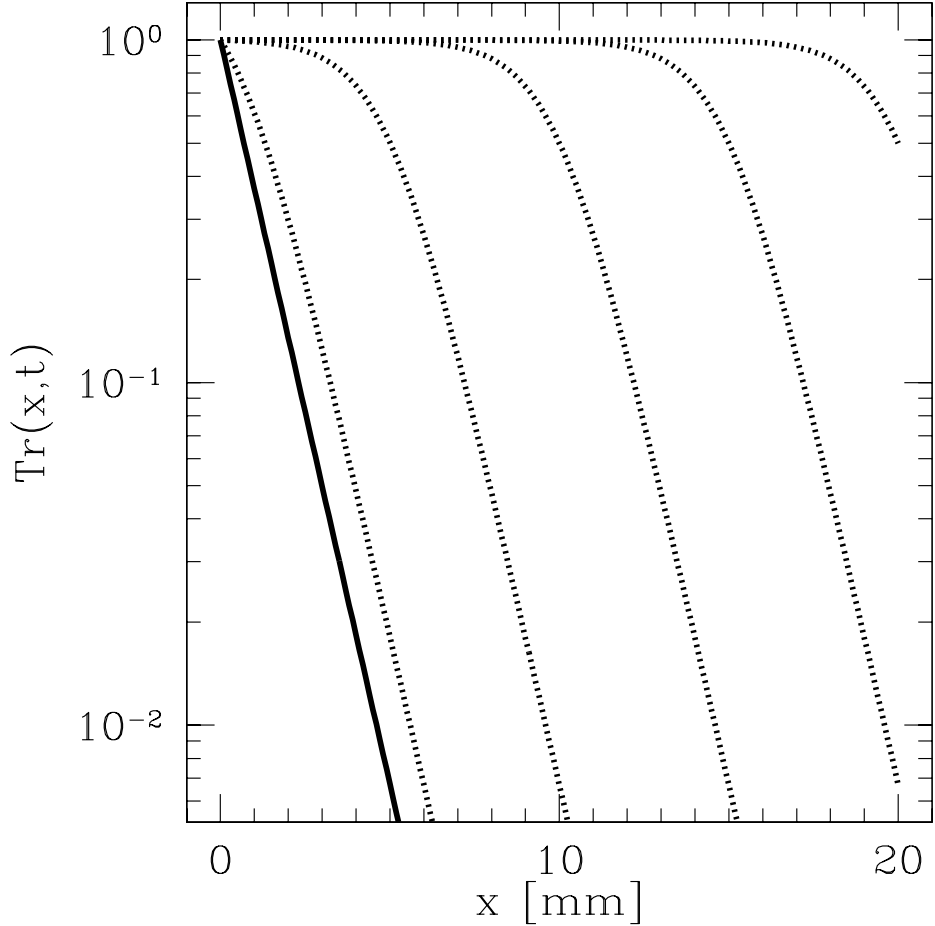}
      \caption{Time evolution of $Tr(x,t)$ as a function of $x$  for both total
        photobleaching (dotted) and photo-invariant polymerization
        (solid), for $\mu_0=1.0$ mm$^{-1}$.  The total photobleaching case is
        shown for dimensionless times of $Kt=1,5,10,15$ and 20 (moving
        from left to right). At long times and
        large $x$, the
        slope of $\ln Tr$ approaches $1/\mu_0$, while for
        $x\rightarrow0$ the slope of $\ln Tr$ approaches $1/\mu_{\infty}$
        [see Eq. (\ref{model-int})].}
   \label{fig:Trsc1and2}
   \end{center}
\end{figure}

The light transmission $Tr$ is similarly exactly calculated as a
function of either ($x,t$) or ($z,t$) as
\begin{eqnarray}
Tr(x,t) &=& \left[{ 1 - \exp(-Kt) + \exp(\mu_0x - Kt)}\right]^{-1};\\
Tr(z,t) &=& \frac{\phi(z)}{1-\exp(-Kt)}
\label{eq-perfecttr}
\end{eqnarray}
This expression reduces to the Beer-Lambert relation, $Tr(x,t
\rightarrow 0^+) = \exp[-\mu_0x]$ for the photopolymerizable
material at short times and $Tr(x,t)$ itself \emph{frontally
propagates} into the medium with increasing time. [$Tr(x,t)$ for
air is unity in our model so that $Tr(x < 0, t) \equiv 1$.] All of
space thus becomes `transparent' to radiation (i.e., $\mu$ = 0) in
the limit of infinite times for total photobleaching, i.e., $
Tr(x, t\rightarrow \infty) = 1$. We plot $Tr(x,t)$ for
representative dimensionless times $Kt$ in Fig.
\ref{fig:Trsc1and2}.

\subsection{Photo-Invariant Polymerization \\
($\mu_0 > 0$ and $\mu_\infty = \mu_0$)}

Another important limit of our FPP model involves the situation in
which the optical attenuation of the polymerized medium is taken
to be unchanged from the pure monomer. This situation is a
reasonable approximation if the monomer is the predominant
component of the photopolymerizable material and if its optical
properties (and density) are insensitive to conversion. In this
\textit{photo-invariant polymerization} case, the conversion
fraction equals
\begin{equation}
\phi(x,t)=1-\exp[-K \exp(-\mu_0x)t]. \label{photoinv-phi}
\end{equation}
As in the previous limiting case, note that this expression
reduces  to Eq. (\ref{eq:phi0}) when $x=0$.
(Curiously, $1 - \phi(x,t)$ is the Gumbel function \cite{gumbel}
of extreme value statistics.) Eq. (\ref{photoinv-phi}) can be
written in the coordinate frame $z$ of the moving front as,
\begin{eqnarray}
\phi(z,t) &=& 1- \exp[-\exp(-\mu_0z)]
\\
z\equiv (x-x_f), & & x_f=\frac{\ln(Kt)}{\mu_0}, \label{photoinv-phim}
\end{eqnarray}
and we have plotted $\phi(z)$ and $x_f$ for this limiting case in
Figs. \ref{fig:phisc1and2} and \ref{fig:xpsc1and2}. We note that
$x_f$ is the position of the inflection of $\phi(x,t)$, and
$\phi=1-e^{-1}\approx 0.632$ at this point. We see from this plot
that $\phi(x,t)$ once again has an invariant sigmoidal shape.

As before, we define the height $h(t)$ of the FPP front by the
condition $\phi(h,t) = \phi_c$:
\begin{equation} \phi_c = 1 - \exp[-K \exp(-\mu_0h)t]
\end{equation}

\noindent and we infer that the height $h(t)$ of the front grows
\textit{logarithmically} with time [see Eqn.
(\ref{photoinv-phim}), and \cite{cabral2004}]

\begin{eqnarray}
h(t, \mu_0, K, \phi_c) &=& \frac{\ln(t/\tau)}{\mu_0}
\label{photoinv-ha}
\\
\tau(K,\phi_c) & \equiv & \frac{\ln[1/(1-\phi_c)]}{K}
\label{photoinv-h}
\end{eqnarray}

\noindent This logarithmic front movement is contrasted with the
linear frontal kinetics of the perfect photobleaching case. The
expression for $h(t)$ in Eq.~(\ref{photoinv-ha}) is restricted to
$t>\tau$ since the solidification front does not form
instantaneously with light exposure, but grows at $x=0$ as
dictated by Eq. (\ref{eq:phi0}). Thus, an induction time $\tau$ is
required for $\phi$ to first approach $\phi_c$ and for the front
to begin propagating. The magnitude of the induction time depends
on the selected threshold $\phi_c$, becoming much larger for $x_f$
as $\phi_c$ approaches $\phi$ at the inflection point, $\phi_f$
[see Eq. (\ref{photoinv-h})]. Notice that the slope of the
$\ln(t)$ factor, describing the growth of $h(t)$ in Eq.
(\ref{photoinv-ha}), depends only on the optical attenuation
$\mu_0$ rather than the rate of reaction and that the intercept
governing the initial front growth is governed by $\tau$, which in
turn depends on the rate constant, optical intensity and $\phi_c$.
Such travelling wavefronts with a logarithmic displacement in time
occur in diverse contexts \cite{majumdar01, majumdar02}. Our
measurements of FPP with a thiol-ene photopolymerizable material
have generally indicated logarithmic front displacement over
appreciable time scales (see Fig. 3 and
\cite{cabral2004,cabral2005}).

The transmission $Tr(x,t)$ does not evolve in time for
photo-invariant polymerization; $Tr(x,t)$ simply decays
exponentially with depth ($x$) according to the Beer-Lambert
relation, $Tr(x,t) = \exp (-\mu_0x)$. This invariance with time is
contrasted in Fig. \ref{fig:Trsc1and2} with the wave-like
propagation of $Tr(x,t)$ in the photobleaching case, corresponding
to the invasion of the polymerizable material of attenuation
$\mu_0$ by an optically transparent medium.

It is important to realize that Eq. (\ref{photoinv-ha}) describes
the initial FPP growth process for an \textit{arbitrary} optical
attenuation of the polymerized material ($\mu_0>0$). Moreover, Eq.
(\ref{photoinv-ha}) describes the long time asymptotic growth
provided that $\mu_0$ is replaced by its non-vanishing counterpart
$\mu_\infty$ for the fully polymerized material. These extremely
useful approximations arise simply because $\mu(x,t)$ is slowly
varying in these short and long time ``fixed-point'' limits. The
crossover between these limiting regimes can be non-trivial and is
addressed below. In many practical instances, however, the time
range is restricted to the initial stage governed by Eq.
(\ref{photoinv-ha}).

\subsection{General FPP solution}

Previous investigations of FPP have relied on numerical solutions
of the governing kinetic equations in comparison to FPP
measurements validating the model. These treatments were
sufficient to demonstrate a good consistency between the model and
experiment \cite{cabral2004,cabral2005}, but many aspects of the
model are difficult to infer in the general case without a full
analytic treatment of the problem.

First, we define the transform variables $\theta= - \ln(1-\phi)$
and $\delta= -\ln(Tr)$. Eqs. (\ref{model-phi}) and
(\ref{model-int}) are then rewritten as
\begin{equation}
  \label{eq:thetadot}
  \frac{\partial\theta}{\partial t}= K e^{-\delta},
\end{equation}
and
\begin{equation}
  \label{eq:deltadx}
  \frac{\partial\delta}{\partial x} = \mu_\infty-(\mu_\infty-\mu_0) e^{-\theta}.
\end{equation}
We now take the $x$-derivative of Eq. (\ref{eq:thetadot}) and the
$t$-derivative of Eq. (\ref{eq:deltadx}) and subtract the
resulting equations obtaining
\begin{equation}
  \label{eq:diff}
  \frac{\partial^2}{\partial x\partial
  t}(\delta-\theta)=\mu_\infty\frac{\partial\theta}{\partial t}.
\end{equation}
This equation  can be integrated directly, yielding
\begin{equation}
  \label{eq:diffint}
    \frac{\partial }{\partial x}(\delta-\theta)=\mu_\infty\theta+c_1,
\end{equation}
where $c_1(x)$ is an arbitrary function of $x$. We now impose the first
of two boundary conditions: namely that at $t=0$, $\phi=0$ for all
$x$ so $\theta(x,0)=0$, while $\delta(x,0)=\mu_0 x$.  This implies
that $c_1(x)=\mu_0$, a constant.   If we now insert
Eq. (\ref{eq:deltadx}) into 
Eq. (\ref{eq:diffint}) we find
\begin{equation}
  \frac{\partial\theta}{\partial x}=
(\mu_\infty-\mu_0)\left(1-e^{-\theta}\right)-\mu_\infty\theta,
\label{eq:dthetadx}
\end{equation}
which again can be integrated. This integration gives
\begin{equation}
  \label{eq:secondint}
x=\frac{1}{\mu_\infty}\int^\theta_{\theta_
0}\frac{d\theta'}{\lambda\left(1-e^{-\theta'}\right)-\theta'},
\end{equation}
where we  define $\lambda\equiv 1-\mu_0/\mu_\infty$ and impose the
second boundary condition $\theta_0 = Kt$ (see Eq. \ref{eq:phi0}.)
Note that $\theta_0$ is the dimensionless time introduced above.
An expression for $\theta$ is obtained by defining the auxiliary
function, $\cal J_\lambda(\theta)$,
\begin{equation}
    \label{eq:defJ}
    {\cal J_\lambda(\theta)}\equiv\int^\theta_1 \frac{d\theta'}{\lambda\left(1-e^{-\theta'}\right)-\theta'}.
\end{equation}
Although ${\cal J_\lambda}(\theta)$ is non-standard, it can be
readily determined as with other, more familiar, special
functions. The existence of an inverse function of ${\cal
J_\lambda}(\theta)$
 is guaranteed if $\lambda<1$, which is assured by the physics of
the problem (since this restriction simply implies $\mu_0>0$).
Insight into ${\cal
  J}_\lambda(\theta)$ is found by noting that for large values of its
argument, ${\cal J}_\lambda(\theta)$ is well approximated by,
\begin{equation}
  \label{eq:jlamapprox_lg}
 {\cal J_\lambda}(\theta)\approx {\cal J_\lambda}(C)+\ln|\lambda-C|-\ln|\lambda-\theta|.
\end{equation}
where $C\gg1$ is a point of expansion. For small values of the
argument we can develop another expansion about $c\ll 1$
\begin{equation}
  \label{eq:jlamapprox_sm}
 {\cal J_\lambda}(\theta)\approx {\cal J_\lambda}(c)+\frac{\ln(c)}{1-\lambda}-\frac{1}{1-\lambda}\ln\theta.
\end{equation}
For much of the range of its arguments,
\[
\left({\cal
J}_\lambda(\theta)-\mathrm{const}\right)\propto\ln(\theta).\]

\noindent We can now rewrite Eq. (\ref{eq:secondint}) as
\begin{equation}
  \label{eq:nearfinal}
  \mu_\infty x= {\cal J_\lambda(\theta)}- {\cal J_\lambda}(Kt).
\end{equation}
Eq. (\ref{eq:nearfinal}) fully solves the problem, since we can
now write $\theta(x,t)$ formally as
\begin{equation}
  \label{eq:final}
    \theta(x,t)= {\cal J_\lambda}^{-1}(\mu_\infty x+ {\cal
      J_\lambda}(Kt));\quad \phi(x,t)=1-e^{-\theta}.
\end{equation}
Note that the dependencies upon $x$ and $t$
are fully separated, implying a functional invariance in the propagation
of the $\phi$ interface's shape. We explore this invariance in detail
below.

We can also solve for $Tr$, using the formal solution to Eq.
(\ref{model-int})
\begin{equation}
  \label{eq:formalI}
  Tr(x,t)=\exp\left\{-\int_0^x dx'\left[\mu_0(1-\phi(x',t))+\mu_\infty\phi(x',t))\right]\right\}.
\end{equation}
Remarkably, this can integrated to fully solve the problem:
\begin{equation}
  \label{eq:finalI}
  Tr(x,t)=\frac{\lambda\phi+\ln(1-\phi)}{\lambda\phi_0+\ln(1-\phi_0)}=
 \frac{\lambda\phi+\ln(1-\phi)}{\lambda (1-e^{-Kt})-Kt}.
\end{equation}
The solutions for the basic measurable variables $\phi(x,t)$ and
$Tr(x,t)$ are now formally complete. Using any simple mathematical
software the above solutions  can be implemented, solved and
plotted.

\subsubsection{Shape of the interface}
Based on our experience with the two limiting cases of total
photobleaching and photoinvariant polymerization, we expect the
interface shape to be sigmoidal. We now analyze the solution in an
effort to determine its general properties, without reference to
particular parameter choices. We know that $\phi(x,t)$ should
increase at any fixed position  as $t$ increases. From Eq.
\ref{eq:dthetadx} we can easily find $\partial\phi/\partial x$ as,
\begin{equation}
  \label{eq:partialphi}
  \frac{\partial\phi}{\partial x}= \mu_\infty(1-\phi)\left[\lambda\phi+\ln(1-\phi)\right].
\end{equation}
Since $\lambda\le1$ and $\mu_\infty>0$, we see that
$\partial\phi/\partial x$ is always less than $0$. This is our
first observation about the shape of the curve: its slope is such
that $\phi$ monotonically decreases as $x$ increases.  Our second
observation comes from Eq. (\ref{eq:phi0}), where we see that
$\phi_0=\phi(0,t)$ rises from $0$ to $1$ as $t$ increases, while
$\phi(x\rightarrow\infty)$ approaches $0$. We also note that since
 $\phi$ monotonically decreases as $x$ increases, $\phi_0$
is the \emph{time-independent} maximum value of $\phi$. This
property derives from the invariance of $\phi(x)$ interface shape
in time (see below).

As for the two limiting cases, the shape can be further examined
by computing the inflection point of $\phi$, e.g. the extremum of
$\partial\phi/\partial x$:
\begin{equation}
  \label{eq:d2phidx2}
  \frac{\partial^2\phi}{\partial x^2}\propto\lambda(1-2\phi_f)-1-\ln(1-\phi_f)=0,
\end{equation}
where $\phi_f$ is the value of $\phi$ at the inflection point. It
is interesting that the value of $\phi_f$ at the inflection point
can be determined by an equation involving elementary functions,
while the determination of the position of this point, $x_f$,
requires the use of our auxiliary function ${\cal J}_\lambda$. If
we desire, we can compute $\theta_f$ using
$\theta_f=-\ln(1-\phi_f)$. Note that for physical values of
$\lambda<1$, there is only one solution to Eq.
(\ref{eq:d2phidx2}). This unique value of $\phi_f$ can exceed the
maximum value of $\phi$, which occurs, as noted above, at
$\phi(0,t)$.  In this case the plot of $\phi(x,t)$ has no
inflection point. Once the induction time  $t> -\ln(1-\phi_f)/K$
has passed, then the inflection point exists for positive values
of $x$.

Thus, we have a detailed picture of the interface profile
characteristics:
\begin{enumerate}
\item  The maximum value of $\phi(x,t)$ at any given time is always at
  $x=0$, and this maximum  value, $\phi_0=\phi(0,t)$,  rises in time
  as   $\phi_0=1-\exp(-Kt)$ .
\item Both $\phi$ and $\partial\phi/\partial x$ approach 0 when $x\rightarrow\infty$.
\item $\partial\phi/\partial x<0$ for all values of $x$, thus $\phi$
  decreases monotonically with increasing $x$.
\item There is a single extreme value of the slope
  $\partial\phi(x,t)/\partial x$. This extremum is found
 when $\phi=\phi_f$ as dictated by Eq. (\ref{eq:d2phidx2}), but only when
 $t> -\ln(1-\phi_f)/K$.
\end{enumerate}
This description outlines precisely the sort of  sigmoidal shape we
expected based on our physical understanding of the system.

\subsubsection{Position of the interface}
We next explore the properties of the above solution for
$\phi(x,t)$ in as much generality as possible.  Eq.
(\ref{eq:nearfinal}) is particularly  illuminating, since it can
be rewritten as
\begin{eqnarray}
  \label{eq:xstar}
  z\equiv x-x^*&=&\frac{1}{\mu_\infty}{\cal J_\lambda}(\theta)-x_0\\
  x^*&=&x_0-\frac{1}{\mu_\infty}{\cal J_\lambda}(Kt),
\end{eqnarray}
where we now see that the shape of the interface is {\sl
invariant} in time, as it was in the limiting cases, and
propagates with the position $x^*(t)$. Indeed, we can invert Eq.
(\ref{eq:xstar}) and write
\begin{equation}
  \label{eq:inverseofz}
\theta(z)=-\ln(1-\phi(z))={\cal J}_\lambda^{-1}(\mu_\infty(z+x_0))
\end{equation}

\begin{figure}[htbpH]
   \begin{center}
      \includegraphics[width=13cm]{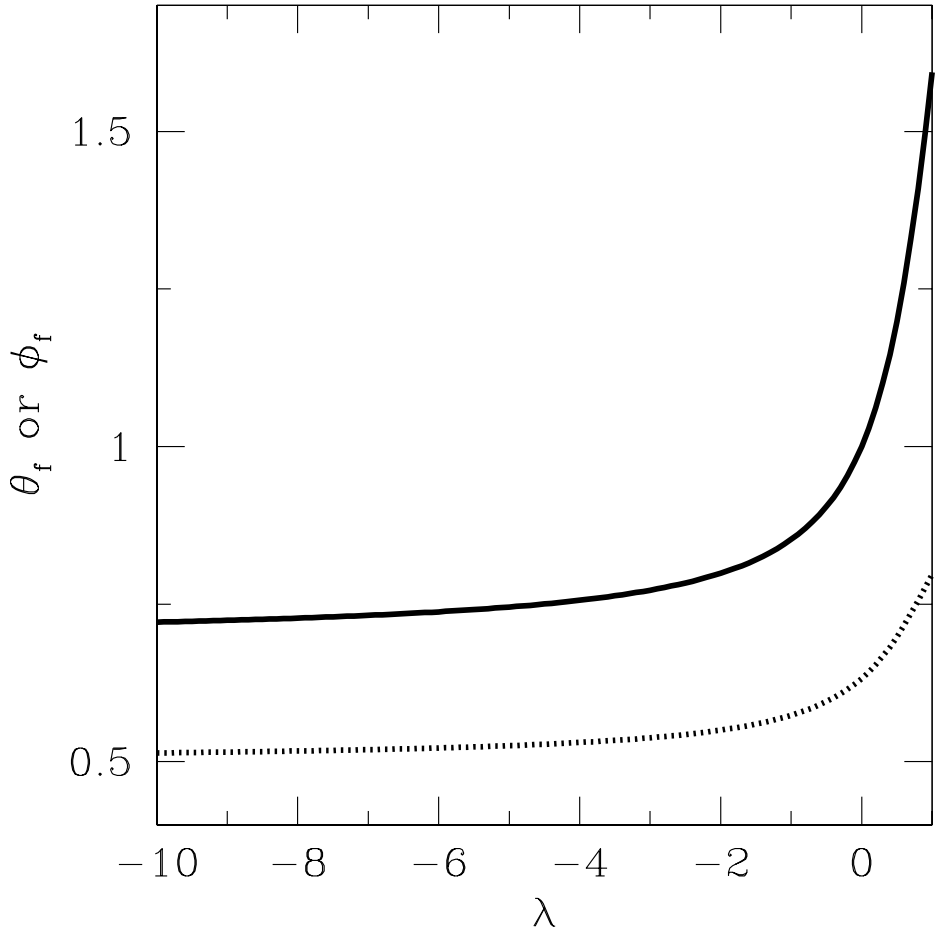}
      \caption{Plot of $\theta_f$ (solid upper curve) and $\phi_f$
        (dotted lower curve) as a function of $\lambda$. In the
        limit $\lambda\rightarrow -\infty$, $\phi_f\rightarrow 0.5$
        and $\theta_f\rightarrow \ln 2$.}
   \label{fig:thetai}
   \end{center}
\end{figure}

While we are free to choose any value of the offset of the
interface position $x_0$, several choices present themselves. One
is the position of the inflection point $x_f$, defined by the
solution to Eq. (\ref{eq:xstar}) with $\theta=\theta_f$, found
from Eq. \ref{eq:d2phidx2}.  If we set $x^*=x_f$ we then have the
formal equations describing the interfacial positions,
\begin{eqnarray}
x_0&=&\frac{1}{\mu_\infty}{\cal J_\lambda}(\theta_f),\\
x_f&=&\frac{1}{\mu_\infty}\left({\cal J_\lambda}(\theta_f)-{\cal J_\lambda}(Kt)\right)
  \label{eq:x0toxi}
\end{eqnarray}
We can get more insight into the properties of this inflection
point by calculating the solution to Eq. (\ref{eq:d2phidx2}) for
all $\lambda\in(-\infty,1]$. Both $\theta_f$ and $\phi_f$ vary
over a fairly narrow range,  as is seen  in Fig. \ref{fig:thetai},
where
 $\phi_f\in[0.5,~0.797]$ and  $\theta_f\in[\ln2,~1.594]$.

As was done in the limiting cases, we can also define the ``height
function'' $h(t)$,  by  choosing a particular value of
$\phi=\phi_c$ which marks the interface position.  With this
choice, and the relation $\theta_c=-\ln(1-\phi_c)$, we then have
the formal expressions
\begin{eqnarray}
  \label{eq:hp}
  x_0&=&\frac{1}{\mu_\infty}{\cal J_\lambda}(\theta_c),\\
h(t) &=&\frac{1}{\mu_\infty}\left({\cal J_\lambda}(\theta_c)-{\cal
    J_\lambda}(Kt)\right).
\end{eqnarray}
The only difference between $x_f$ and $h$ is the fixed `offset'
\begin{equation}
  \label{eq:hxidiff}
  h(t) -x_f=\frac{1}{\mu_\infty}\left({\cal J_\lambda}(\theta_c)-{\cal
    J_\lambda}(\theta_f)\right).
\end{equation}

\subsubsection{Induction time}
In our study of the limiting cases, we found an induction time
when $\phi(x,t)$ first exceeded $\phi_f$ (at the inflection point)
or $\phi_c$ (the physically selected interface position).  In
general, regardless of what convention we choose for the interface
position, the induction time will be simply the solution to
$\phi_0=\phi^*$, where $\phi^*$ is the value of $\phi$ at the
interface $\phi_f$ or $\phi_c$. Using Eq. (\ref{eq:phi0}) this is
simply
\begin{equation}
  \label{eq:induction}
  \tau=\frac{-\ln(1-\phi^*)}{K}
\end{equation}

Because of this induction time, and the different values of
$\phi^*$ used in our definitions of $x_f$ and $h$, these functions
can actually behave quite differently at early times. Typically we
select $\phi_c\ll 1$, and thus, for this case, the induction time
will be relatively short on experimental time scales, $\tau\approx
\phi_c/K$. On the other hand, our computation of the range of
$\phi_f\in(0.5,0.797)$ implies a range in $\tau\in(0.693,1.594)$.
These values of $\tau$ are between 35 and 80 times larger than
induction times established using a typical choice of
$\phi_c=0.02$.

\subsubsection{Approximations to the front position}

It is useful to obtain approximate expressions for Eq.
(\ref{eq:xstar}).  At early times we can develop an approximation
solely for $h$, since $x_f$ is undefined at early times. Using Eq.
(\ref{eq:jlamapprox_sm}) we obtain the explicit estimate
\begin{equation}
  \label{eq:happrox}
  h(t)\approx \frac{1}{\mu_0}\ln\left(\frac{Kt}{\theta_c}\right);\ \theta_c<Kt\ll1
\end{equation}
Thus, an early time log-linear plot of $h(t)$ will yield a slope
of $1/\mu_0$.  Note that this expression is {\sl exact} for the
case of photo-invariant polymerization, as comparison with Eq.
(\ref{photoinv-h}) reveals.

At long times, we can develop a general expression for an
approximate form to $x^*$  using Eq. (\ref{eq:jlamapprox_lg}).
Thus, we introduce an expansion for the limit $C\gg1$,
\begin{equation}
  \label{eq:xpapprox}
  x^*\approx x_0-c_1+\frac{1}{\mu_\infty}\ln|\lambda-Kt|
\end{equation}
where $c_1=[{\cal J}_\lambda(C)+\ln|\lambda-C|]/\mu_\infty$.

We recall, however, the limiting case where $\mu_\infty=0$ (total
photobleaching) yields,
\begin{eqnarray}  x^*&=&x_0+\frac{1}{\mu_0}
  \left[Kt+\ln\left(1-e^{-Kt}\right)\right],\\
x_0&=&\frac{1}{\mu_0}\ln\left(\frac{1}{\phi^*}-1\right),\end{eqnarray}
which has a {\sl linear} $x^*\propto t$ behavior at long times.
This seems quite different from the logarithmic behavior given
above for the general expression.  How can this be understood?  In
the limit $\mu_\infty\rightarrow 0$, we  have that
$\lambda\rightarrow -\infty$.  For any non-zero value of
$\mu_\infty$ the logarithmic behavior the approximate form {\sl
must} dominate at long times. However there will always be an
intermediate time (perhaps a very long time if $|\lambda|$ is
large!) when $Kt\ll |\lambda|$, and in this case we can expand the
$\ln|\lambda-Kt|\approx \ln|\lambda|-Kt/\lambda$ to obtain linear
behavior.
\begin{equation}
  x^*\approx x_0-c_1+\frac{Kt}{\mu_0
    -\mu_\infty}+\frac{\ln|\lambda|}{\mu_\infty};\ 1\ll Kt\ll |\lambda|
\end{equation}

Now that we have a ``general'' solution to our kinetic equations,
we examine the specific cases of  partial photodarkening and
partial photobleaching.

\subsection{Illustration of general solution: partial photobleaching vs. photodarkening }

The limits of perfect photobleaching and photo-invariant
polymerization are ideals that only approximately arise in
practice. In general, the optical attenuation of the polymerizable
material is always greater than zero and can either increase or
decrease upon conversion.

It is possible that the reactive products generated by the
photoinitiator or the polymerization of the monomer increase the
optical attenuation so that the polymerized material becomes
increasingly opaque to radiation with increasing time: partial
photodarkening ($\mu_\infty > \mu_0$). We find that this is a
common situation in our FPP measurements, regardless of the
presence of nanoparticles, or temperature variations
\cite{cabral2004,cabral2005}. For this case we choose the
realistic model parameters: $\mu_0$ = 1 mm$^{-1}$, $\mu_\infty =
5.0$ mm$^{-1}$, $K$= 1 s$^{-1}$. As mentioned before, we select
the representative value for $\phi_c$ = 0.02.

\begin{figure}[htbpH]
   \begin{center}
      \includegraphics[width=13cm]{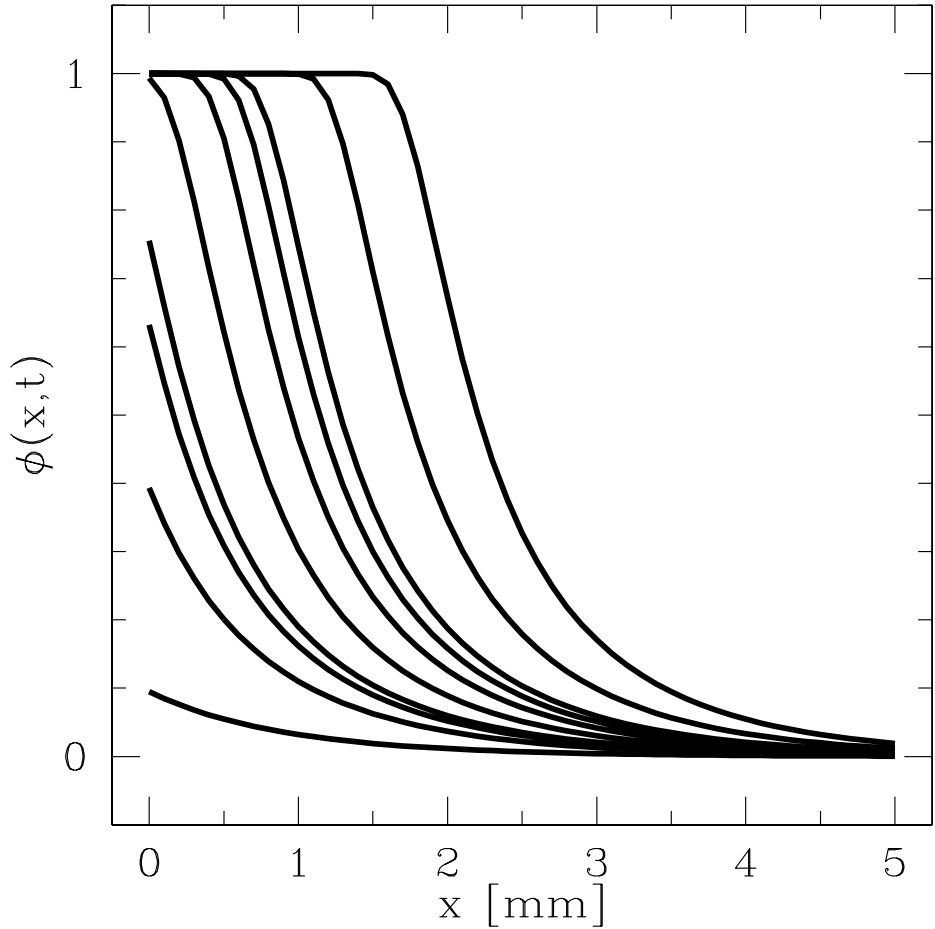}
      \caption{Evolution of the conversion $\phi$ with time for
        partial photodarkening (parameters in text), plotted (going up
        and leftward) at $Kt=$0.1, 0.5, 1.0, 1.40897, 5,
        20, 50, 100, 1000, and 10000.}
   \label{fig:phi}
   \end{center}
\end{figure}

\begin{figure}[htbpH]
   \begin{center}
      \includegraphics[width=13cm]{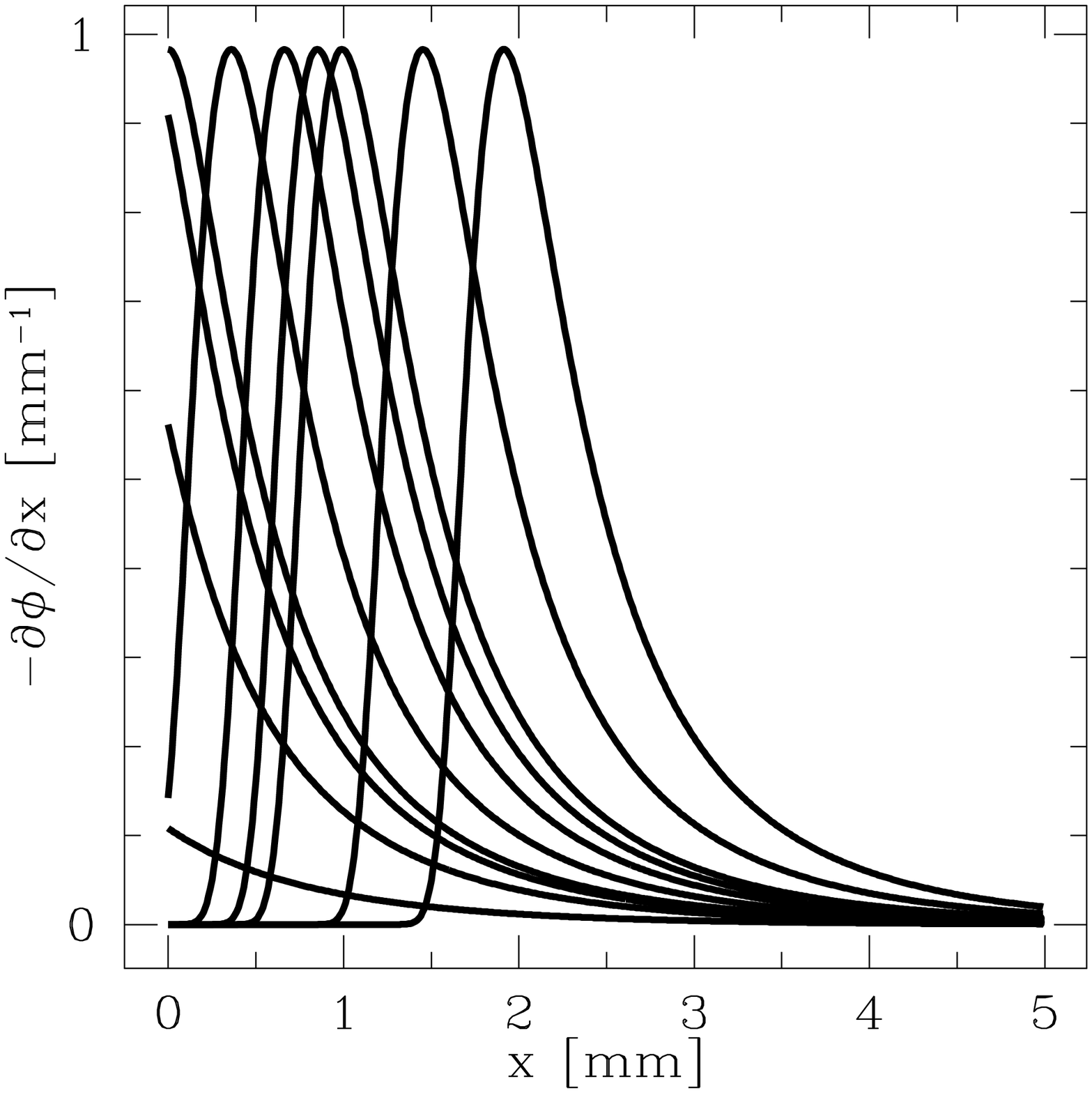}
      \caption{Evolution of $-\partial\phi/\partial x$ in time for
        partial photodarkening, shown (going up
        and left to right) at $Kt=$0.1, 0.5,
        1.0, 1.40897, 5, 20, 50, 100, 1000, and 10000.}
   \label{fig:dphidx}
   \end{center}
\end{figure}

The spatio-temporal variation of the conversion fraction
$\phi(x,t)$ is shown in Fig. \ref{fig:phi} and its the derivative
$-\partial \phi(x,t)/\partial x$ is shown in Fig.
\ref{fig:dphidx}. (Since the slope is negative definite, we plot
its magnitude $-\partial\phi/\partial x$). We see the development
of a well defined advancing front as in the perfect photobleaching
and photo-invariant limits, discussed above. We compare these
results with other choices of the parameters below.

In contrast to photodarkening, we also consider the case where
$\mu_\infty$ is small: partial photobleaching ($\mu_\infty <
\mu_0$). Specifically,  we keep all other parameters the same but
reduce $\mu_\infty$ by a factor of 10. Thus, $\mu_\infty=0.5$
mm$^{-1}$, implying $\lambda=-1$. The behavior of this system
should then be somewhere between the partial photoinvariant case
and the total photobleaching limit.

Note that the frontal kinetics of FPP is specified by only 4 basic
model parameters in the framework of our model: $\mu_0$,
$\mu_\infty$, $K$ and $\phi_c$. The attenuation coefficients can
be determined independently with a set of $Tr$ vs. thickness
experiments of the neat and fully polymerized material (Fig.
\ref{fig-exp}). $K$ may be determined by the time (or dose)
\textit{dependence} of the $Tr$, for various thicknesses. Finally,
the solidification conversion threshold $\phi_c$ is obtained by
fitting measurements of \textit{height} as a function of
\textit{dose} to our theory.

We next consider a comparative analysis of the FPP front cases.
The extent of polymerization conversion fraction $\phi$ propagates
as a shape invariant waveform, after an induction period. The time
evolution of $\phi$ for partial photodarkening is illustrated in
Fig. \ref{fig:phi} with parameters $\lambda=0.8$ and
$\mu_\infty=5.0$. We find from Eq. \ref{eq:d2phidx2} that
$\phi_f=0.755605$, and therefore
$\theta_f=-\ln(1-0.755605)=1.40897$. Accordingly, the shape of
$\partial\phi/\partial x$, plotted in Fig. \ref{fig:dphidx}, is
invariant for $Kt>1.40897$ and simply propagates to the right as
$t$ increases. For the partial photobleaching case we find
$\theta_f=0.852606$ and $\phi_f=0.573697$. This shape invariance
is best understood and appreciated by transforming $\phi$ into the
moving coordinate $z$ of the front (as in Fig.
\ref{fig:phisc1and2}), which is shown below. First, however, we
consider the time dependence of the position of the front. As
before, the location of the peak in $\partial\phi/\partial x$
defines $x=x_f(t)$ [Eq. (\ref{eq:x0toxi})]. We now see why $x_f$
is also a suitable alternative choice for the position of the FPP
front (particularly if optical methods are used to locate the
interface experimentally). Evidently, the peak height and shape of
$\partial\phi/\partial x$ are invariant after the peak first
appears at $Kt>\theta_f$.

\begin{figure}[htbpH]
   \begin{center}
      \includegraphics[width=13cm]{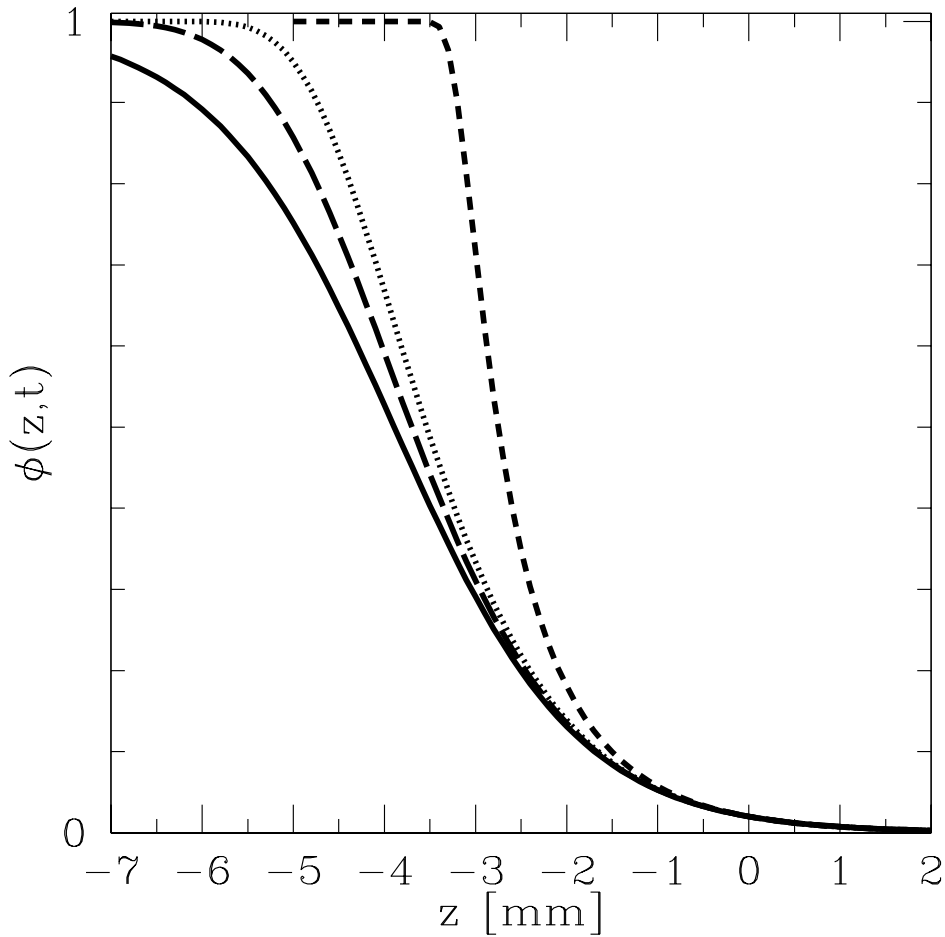}
      \caption{Conversion $\phi(z)$ as a function of the coordinate in the moving frame of the front $z$ for 4 different
        cases:  total photobleaching (solid, $\lambda=-\infty$),
        partial photobleaching (long-dash,
        $\lambda=-1.0$),  photoinvariant (dotted, $\lambda=0$), and partial
        photodarkening
        (short-dash, $\lambda=0.8$). The plots were chosen so that they
        intersect at $\phi=\phi_c$.  The profiles are \emph{time-invariant}.}
   \label{fig:phivxhat}
   \end{center}
\end{figure}

The time-invariant nature of the front propagation of $\phi$ in
the moving frame is illustrated in Fig. \ref{fig:phivxhat}. We
observe that the $\phi(z)$ profiles are sigmoidal and independent
of time when plotted with respect to the transformed variable
$z=x-h(t)$. All curves intersect when $\phi=\phi_c$, explaining
the overlap at low values of $\phi(z)$.

\begin{figure}[htbpH]
   \begin{center}
      \includegraphics[width=13cm]{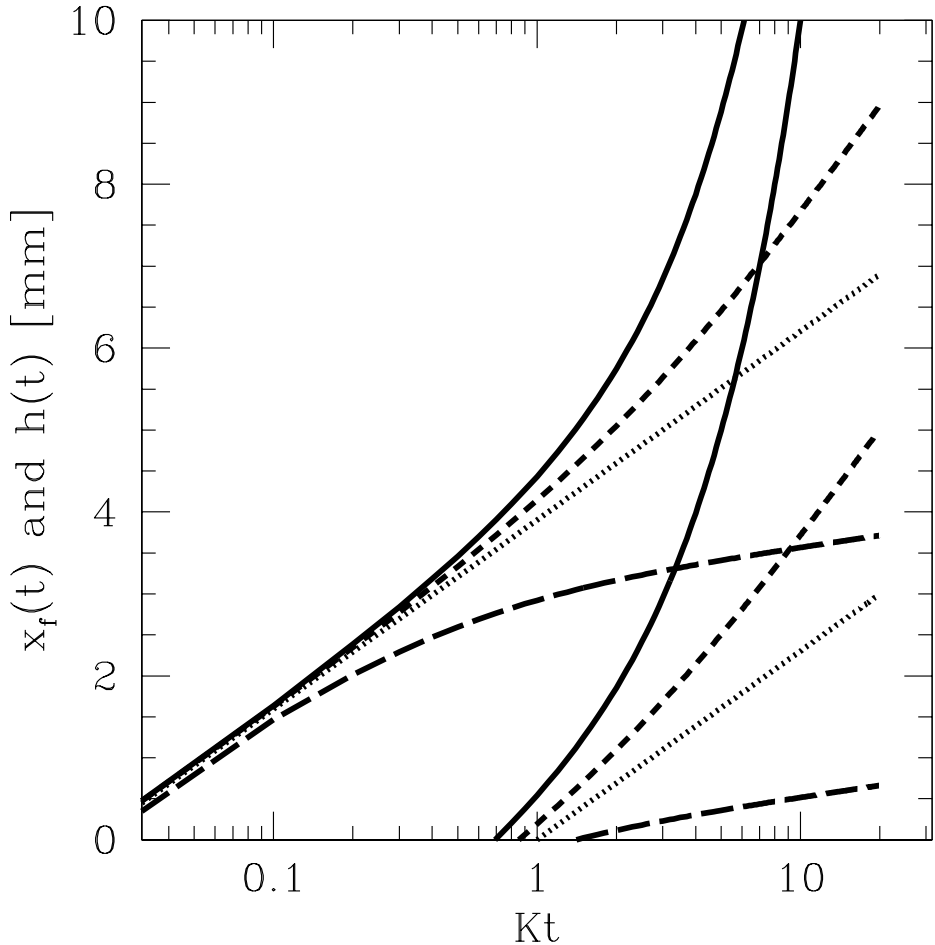}
      \caption{Front position $x_f$ (emerging from the $Kt$-axis near
        $Kt\approx 1$) and $h$ (emerging as  $Kt\rightarrow0$) as a function of
        $Kt$. We show all 4 cases: total photobleaching (solid, $\lambda=-\infty$),
        partial photobleaching (long-dash, $\lambda=-1.0$),
        photoinvariant (dotted, $\lambda=0$), and partial
        photodarkening (short-dash, $\lambda=0.8$). The interface appears after its induction time
        $Kt=-\ln(1-\phi^*)$, where $\phi^*=\phi_c=0.02$ for the plots
        of $h$ while $\phi^*=\phi_f$ for the case of $x_f$.}
   \label{fig:xandh}
   \end{center}
\end{figure}

Since $x_f(t)$ and $h(t)$  are both important measures of FPP
frontal kinetics, we compute these observable quantities in Fig.
\ref{fig:xandh} for all four cases: total photobleaching (solid,
$\lambda=-\infty$), partial photobleaching (long-dash,
$\lambda=-1.0$), photoinvariant (dotted, $\lambda=0$), and partial
photodarkening (short-dash, $\lambda=0.8$). In all cases, the
interface evidently appears after its (dimensionless) induction
time $Kt=-\ln(1-\phi^*)$, where $\phi^*=\phi_c=0.02$ for the
height $h$ (group emerging near $Kt\rightarrow0$), while
$\phi^*=\phi_f$ for the inflection 
point front position $x_f$ (group emerging near $Kt\approx 1$), where
$\phi_f$ is found 
from Eq. (\ref{eq:d2phidx2}). Note that the vertical offset
between $x_f$ and $h$  is the constant $x_0$ dictated by Eq.
(\ref{eq:x0toxi}). All the examples shown reach $x_f,\ h\propto
\ln Kt$ at late times (near where $Kt>|\lambda|$), except for
total photobleaching, which remains in linear growth kinetics at
late times.

As in the total photobleaching case, we see that the FPP front
position (as defined by the inflection point) is insensitive to
crossover effects since this feature develops at late times (see
Fig. \ref{fig:xandh}). The displacement in time is logarithmic
after a short induction time. The front position $h(t)\equiv x
(\phi=\phi_c)$, as defined by a `critical' conversion (here,
$\phi_c=0.02$), does exhibit a noticeable crossover. As
anticipated from Eq. (5), the front position $h(t)$ moves
logarithmically at `short' times where $\bar{\mu}(x,t\rightarrow
0) \approx \mu_0$ and crosses over to a slope determined by
$\bar{\mu}(x,t\rightarrow \infty) \approx \mu_\infty$,
respectively, as the monomer interconverts to a polymerized
network. In the partial photodarkening case, the front moves
faster initially ($ \propto 1/\mu_0$) and slows down ($ \propto
1/\mu_\infty$) at later times. The situation occurs in the case of
partial photobleaching.

\begin{figure}[htbpH]
   \begin{center}
      \includegraphics[width=13cm]{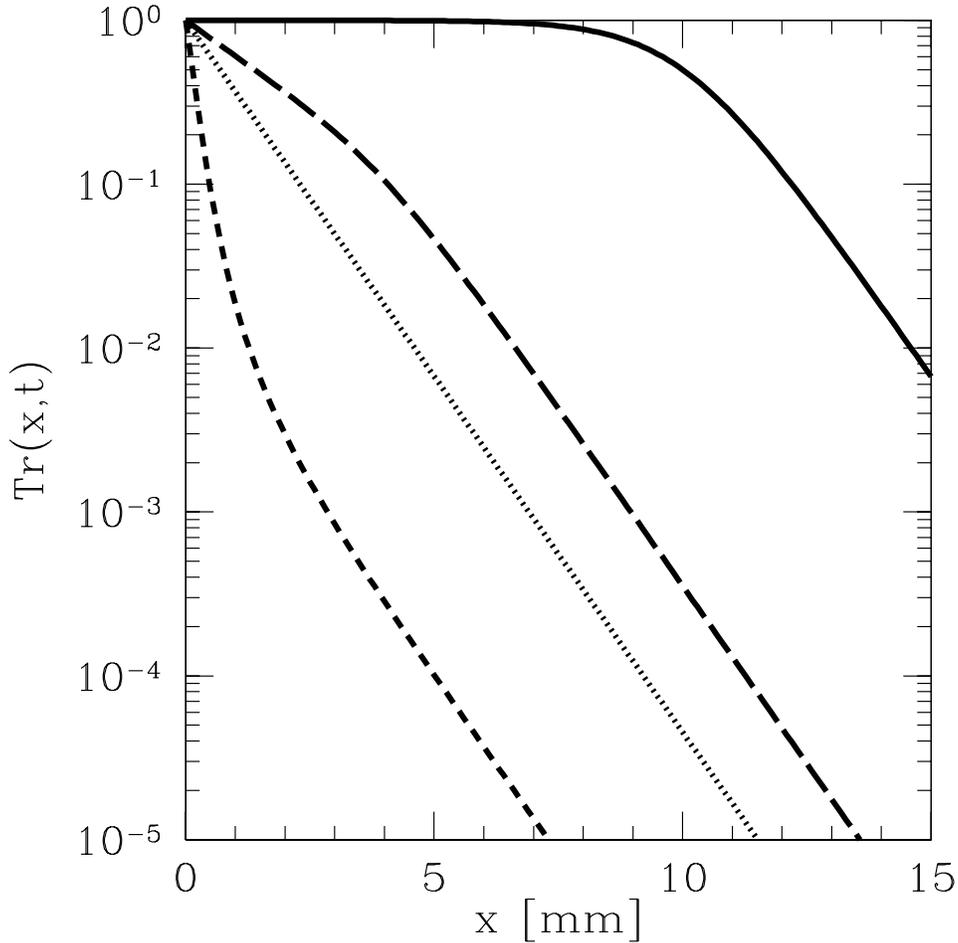}
      \caption{Transmission $Tr(x,t)$ as a function of position at the late
        time of $Kt=10.0$, for 4 different
        cases:  total photobleaching (solid, $\lambda=-\infty$),
        partial photobleaching (long-dash, $\lambda=-1.0$), photoinvariant (dotted, $\lambda=0$, and partial
        photodarkening
        (short-dash, $\lambda=0.8$). The slopes exhibit the expected crossover
        from $1/\mu_0$ to $1/\mu_\infty$. The frontal character of $Tr(x,t)$ is illustrated in Fig. \ref{fig:Trsc1and2} and Figs. 4 and 5 of \cite{cabral2005}.}
   \label{fig:iplot}
   \end{center}
\end{figure}

The evolution of the light intensity is sensitive to the evolution
of the optical attenuation and is thus particularly interesting
and informative about the nature of the front development. The
transmission $Tr(x,t)$ (dimensionless intensity profile) as a
function of depth for various curing times is plotted in Fig.
\ref{fig:iplot} for both sets of parameters. The initial profile
is simply $Tr= e^{-\mu_0 x}$, and it decreases in the manner given
by Eq. (\ref{eq:finalI}). In the short and long time limits, we
see that the usual Beer-Lambert law holds and the intensity decays
exponentially in $x$, with attenuation coefficients $\mu_0$ and
$\mu_\infty$, respectively. At intermediate times, there is a
crossover between these two asymptotic regimes. Note that an
attempt to fit experimental transmission results with the simple
Beer-Lambert law would result in an unphysical ($\neq$ 1)
intercept for infinitely thin films, symptomatic of the necessity
of accounting for the variation in $\mu$ in the course of
photopolymerization. This is how we first recognized the
importance of partial photodarkening in our former measurements
\cite{cabral2004,cabral2005}.

Figures \ref{fig:phivxhat}--\ref{fig:iplot} summarize our findings
for the conversion $\phi(z)$ and light attenuation $Tr(x,t)$
profiles, frontal kinetics (using both inflection $x_p$ and height
$h$ criteria) for the four cases illustrated: total and partial
photobleaching, photoinvariant and partial photodarkening
polymerization. We see from this comparative discussion that,
while the properties of polymerization front propagation in the
unpolymerized material are general, the shape of the fronts $\phi$
and $Tr$ and the time development of the front position (linear
and logarithmic, induction time) depends on the evolution of the
optical attenuation upon polymerization.

\section{Conclusions}

We have exactly solved a model frontal photopolymerization (FPP)
that directly addresses the kinetics of the growth front position
and the change in optical attenuation in time under general
circumstances. This model involves an order parameter $\phi (x,t)$
describing the extent of conversion of monomer to polymer (solid)
and the extent of light attenuation, $Tr(x,t)$. Many aspects of
the photopolymerization process derive from the changing character
of the optical attenuation $\mu$ in the course of PM exposure to
light, and we illustrate how this effect can lead to significant
changes in the kinetics of front propagation.

The optical attenuation of the photopolymerizable material leads to
non-uniformity in the extent of polymerization. Solidification
develops first at the boundary when the polymer conversion becomes
sufficiently high and then a front of solidification invades the
photopolymerizable material in the form of a wave. We find that the
interface between the solid and liquid is described by a
polymerization density profile $\phi(z)$ whose shape is invariant in
time. The time dependence of the front movement and the shape of
$\phi(z)$ depend on the change of the optical attenuation accompanying
polymerization. The position of the front is established  using one of
two methods: by specification of a critical value $\phi_c$ for which
solidification occurs (a convenient definition for photolithography
where the liquid material is simply washed away after photo exposure)
or by determination of the inflection point in $\phi(z)$. We find that
the initial frontal 
growth kinetics are logarithmic in time, governed by the optical
properties of the unconverted material and are followed by a transient
crossover. Front displacement in this crossover regime is complex,
as it depends on whether conversion decreases or increases the optical
attenuation. At long times, the front displacement becomes universally
logarithmic in time (excluding the case of ``perfect photobleaching''
where the optical attenuation after UV exposure exactly vanishes and
fronts propagate linearly in time), but it may take an (impractically)
long time for this asymptotic behavior to be reached. Many of the
asymptotic properties of the general case of evolving optical
attenuation that we describe in our model are
captured in a simplified model in which the optical attenuation is
assumed to be a positive, non-vanishing constant: photo-invariant
polymerization.

Our general treatment of photopolymerization has been found to
quantitatively describe frontal growth in both neat
\cite{cabral2004} and nanoparticle filled \cite{cabral2005}
photopolymerizable materials (thiol-ene copolymers) and to capture
the effect of temperature (through a single rate parameter)
\cite{cabral2005}. This description provides a predictive
framework for controlling the spatial dimension of
photopolymerizable materials for microfluidics and other
applications, where the rapid microfabrication of solid structures
is required.

\section{Acknowledgments}
Support from the NIST Combinatorial Methods Center (NCMC) is
greatly appreciated.


\begin{thebibliography}{38}
\expandafter\ifx\csname natexlab\endcsname\relax\def\natexlab#1{#1}\fi
\expandafter\ifx\csname bibnamefont\endcsname\relax
  \def\bibnamefont#1{#1}\fi
\expandafter\ifx\csname bibfnamefont\endcsname\relax
  \def\bibfnamefont#1{#1}\fi
\expandafter\ifx\csname citenamefont\endcsname\relax
  \def\citenamefont#1{#1}\fi
\expandafter\ifx\csname url\endcsname\relax
  \def\url#1{\texttt{#1}}\fi
\expandafter\ifx\csname urlprefix\endcsname\relax\def\urlprefix{URL }\fi
\providecommand{\bibinfo}[2]{#2}
\providecommand{\eprint}[2][]{\url{#2}}

\bibitem[{\citenamefont{Odian}(1991)}]{odian}
\bibinfo{author}{\bibfnamefont{G.}~\bibnamefont{Odian}},
  \emph{\bibinfo{title}{Principles of Polymerization}}
  (\bibinfo{publisher}{John Wiley \& Sons: New York}, \bibinfo{year}{1991}).

\bibitem[{\citenamefont{Fouassier}(1995)}]{Fouassier}
\bibinfo{author}{\bibfnamefont{J.-P.} \bibnamefont{Fouassier}},
  \emph{\bibinfo{title}{Photoinitiation, Photopolymerization, and Photocuring}}
  (\bibinfo{publisher}{Hanser / Gardner Publications: Cincinnati, OH},
  \bibinfo{year}{1995}).

\bibitem[{\citenamefont{Fouassier and Rabek}(1993)}]{fourab}
\bibinfo{author}{\bibfnamefont{J.-P.} \bibnamefont{Fouassier}}
  \bibnamefont{and} \bibinfo{author}{\bibfnamefont{J.~F.} \bibnamefont{Rabek}},
  \emph{\bibinfo{title}{Radiation Curing in Polymer Science and Technology}}
  (\bibinfo{publisher}{Elsevier Applied Science: London},
  \bibinfo{year}{1993}).

\bibitem[{\citenamefont{Decker}(1998)}]{decker1}
\bibinfo{author}{\bibfnamefont{C.}~\bibnamefont{Decker}},
  \bibinfo{journal}{Polymer Int.} \textbf{\bibinfo{volume}{45}},
  \bibinfo{pages}{133} (\bibinfo{year}{1998}).

\bibitem[{\citenamefont{Decker}(2002)}]{decker2}
\bibinfo{author}{\bibfnamefont{C.}~\bibnamefont{Decker}},
  \bibinfo{journal}{Polymer Int.} \textbf{\bibinfo{volume}{51}},
  \bibinfo{pages}{1141} (\bibinfo{year}{2002}).

\bibitem[{\citenamefont{Harrison et~al.}(2004)\citenamefont{Harrison, Cabral,
  Stafford, Karim, and Amis}}]{harrison}
\bibinfo{author}{\bibfnamefont{C.}~\bibnamefont{Harrison}},
  \bibinfo{author}{\bibfnamefont{J.~T.} \bibnamefont{Cabral}},
  \bibinfo{author}{\bibfnamefont{C.}~\bibnamefont{Stafford}},
  \bibinfo{author}{\bibfnamefont{A.}~\bibnamefont{Karim}}, \bibnamefont{and}
  \bibinfo{author}{\bibfnamefont{E.~J.} \bibnamefont{Amis}},
  \bibinfo{journal}{J. Microeng. Micromach.} \textbf{\bibinfo{volume}{14}},
  \bibinfo{pages}{153} (\bibinfo{year}{2004}).

\bibitem[{\citenamefont{Cabral et~al.}(2004)\citenamefont{Cabral, Hudson,
  Harrison, and Douglas}}]{cabral2004}
\bibinfo{author}{\bibfnamefont{J.~T.} \bibnamefont{Cabral}},
  \bibinfo{author}{\bibfnamefont{S.~D.} \bibnamefont{Hudson}},
  \bibinfo{author}{\bibfnamefont{C.}~\bibnamefont{Harrison}}, \bibnamefont{and}
  \bibinfo{author}{\bibfnamefont{J.~F.} \bibnamefont{Douglas}},
  \bibinfo{journal}{Langmuir} \textbf{\bibinfo{volume}{20}},
  \bibinfo{pages}{10020} (\bibinfo{year}{2004}).

\bibitem[{\citenamefont{Wu et~al.}(2004)\citenamefont{Wu, Mei, Cabral, Xu, and
  Beers}}]{tao}
\bibinfo{author}{\bibfnamefont{T.}~\bibnamefont{Wu}},
  \bibinfo{author}{\bibfnamefont{Y.}~\bibnamefont{Mei}},
  \bibinfo{author}{\bibfnamefont{J.~T.} \bibnamefont{Cabral}},
  \bibinfo{author}{\bibfnamefont{C.}~\bibnamefont{Xu}}, \bibnamefont{and}
  \bibinfo{author}{\bibfnamefont{K.~L.} \bibnamefont{Beers}},
  \bibinfo{journal}{J. Am. Chem. Soc.} \textbf{\bibinfo{volume}{126}},
  \bibinfo{pages}{9880} (\bibinfo{year}{2004}).

\bibitem[{\citenamefont{Cygan et~al.}(2005)\citenamefont{Cygan, Cabral, Beers,
  and Amis}}]{cygan}
\bibinfo{author}{\bibfnamefont{Z.~T.} \bibnamefont{Cygan}},
  \bibinfo{author}{\bibfnamefont{J.~T.} \bibnamefont{Cabral}},
  \bibinfo{author}{\bibfnamefont{K.~L.} \bibnamefont{Beers}}, \bibnamefont{and}
  \bibinfo{author}{\bibfnamefont{E.~J.} \bibnamefont{Amis}},
  \bibinfo{journal}{Langmuir}  (\bibinfo{year}{2005}).

\bibitem[{\citenamefont{Hudson et~al.}(2004)\citenamefont{Hudson, Cabral,
  Goodrum, Beers, and Amis}}]{hudson}
\bibinfo{author}{\bibfnamefont{S.~D.} \bibnamefont{Hudson}},
  \bibinfo{author}{\bibfnamefont{J.~T.} \bibnamefont{Cabral}},
  \bibinfo{author}{\bibfnamefont{W.}~\bibnamefont{Goodrum}},
  \bibinfo{author}{\bibfnamefont{K.~L.} \bibnamefont{Beers}}, \bibnamefont{and}
  \bibinfo{author}{\bibfnamefont{E.~J.} \bibnamefont{Amis}},
  \bibinfo{journal}{Appl. Phys. Lett.}  (\bibinfo{year}{2004}).

\bibitem[{\citenamefont{Khan and Pojman}(1996)}]{pojman1}
\bibinfo{author}{\bibfnamefont{A.~M.} \bibnamefont{Khan}} \bibnamefont{and}
  \bibinfo{author}{\bibfnamefont{J.~A.} \bibnamefont{Pojman}},
  \bibinfo{journal}{Trends Polym. Sci.} \textbf{\bibinfo{volume}{4}},
  \bibinfo{pages}{253} (\bibinfo{year}{1996}).

\bibitem[{\citenamefont{Pojman et~al.}(1996)\citenamefont{Pojman, Ilyashenko,
  and Khan}}]{pojman2}
\bibinfo{author}{\bibfnamefont{J.~A.} \bibnamefont{Pojman}},
  \bibinfo{author}{\bibfnamefont{V.~M.} \bibnamefont{Ilyashenko}},
  \bibnamefont{and} \bibinfo{author}{\bibfnamefont{A.~M.} \bibnamefont{Khan}},
  \bibinfo{journal}{J. Chem. Soc., Faraday Trans.}
  \textbf{\bibinfo{volume}{92}}, \bibinfo{pages}{2825} (\bibinfo{year}{1996}).

\bibitem[{\citenamefont{Lewis et~al.}(2004)\citenamefont{Lewis, DeBisschop,
  Pojman, and Volpert}}]{pojman3}
\bibinfo{author}{\bibfnamefont{L.~L.} \bibnamefont{Lewis}},
  \bibinfo{author}{\bibfnamefont{C.~A.} \bibnamefont{DeBisschop}},
  \bibinfo{author}{\bibfnamefont{J.~A.} \bibnamefont{Pojman}},
  \bibnamefont{and} \bibinfo{author}{\bibfnamefont{V.~A.}
  \bibnamefont{Volpert}}, in \emph{\bibinfo{booktitle}{Nonlinear Dynamics In
  Polymeric Systems}} (\bibinfo{publisher}{Eds. J. A. Pojman, Q.
  Tran-Cong-Miyata Q., pp. 169, ACS Symposium Series 869},
  \bibinfo{year}{2004}).

\bibitem[{\citenamefont{Cabral and Douglas}(2005)}]{cabral2005}
\bibinfo{author}{\bibfnamefont{J.~T.} \bibnamefont{Cabral}} \bibnamefont{and}
  \bibinfo{author}{\bibfnamefont{J.~F.} \bibnamefont{Douglas}},
  \bibinfo{journal}{Polymer}  (\bibinfo{year}{2005}).

\bibitem[{dis()}]{disclaimer}
\emph{\bibinfo{title}{Certain commercial equipment, instruments, or materials
  are identified in this paper in order to specify the experimental procedure
  adequately. such identification is not intended to imply recommendation or
  endorsement by the national institute of standards and technology, nor is it
  intended to imply that the materials or equipment identified are necessarily
  the best available for the purpose.}}

\bibitem[{\citenamefont{Jacobine}(1993)}]{jacobine}
\bibinfo{author}{\bibfnamefont{A.~F.} \bibnamefont{Jacobine}}, in
  \emph{\bibinfo{booktitle}{Radiation Curing in Polymer Science and
  Technology}} (\bibinfo{publisher}{Eds. J.-P. Fouassier and J. F. Rabek, Vol.
  3, Chapter 7, pp. 171, Elsevier Applied Science: London},
  \bibinfo{year}{1993}).

\bibitem[{\citenamefont{Cramer et~al.}(2002)\citenamefont{Cramer, Scott, and
  Bowman}}]{Cramer02}
\bibinfo{author}{\bibfnamefont{N.~B.} \bibnamefont{Cramer}},
  \bibinfo{author}{\bibfnamefont{J.~P.} \bibnamefont{Scott}}, \bibnamefont{and}
  \bibinfo{author}{\bibfnamefont{C.~N.} \bibnamefont{Bowman}},
  \bibinfo{journal}{Macromolecules} \textbf{\bibinfo{volume}{35}},
  \bibinfo{pages}{5361} (\bibinfo{year}{2002}).

\bibitem[{\citenamefont{Cramer et~al.}(2003)\citenamefont{Cramer, Davies,
  O'Brien, and Bowman}}]{Cramer03a}
\bibinfo{author}{\bibfnamefont{N.~B.} \bibnamefont{Cramer}},
  \bibinfo{author}{\bibfnamefont{T.}~\bibnamefont{Davies}},
  \bibinfo{author}{\bibfnamefont{A.~K.} \bibnamefont{O'Brien}},
  \bibnamefont{and} \bibinfo{author}{\bibfnamefont{C.~N.}
  \bibnamefont{Bowman}}, \bibinfo{journal}{Macromolecules}
  \textbf{\bibinfo{volume}{36}}, \bibinfo{pages}{4631} (\bibinfo{year}{2003}).

\bibitem[{\citenamefont{Reddy et~al.}(2003)\citenamefont{Reddy, Cramer, Cross,
  Raj, and Bowman}}]{reddy}
\bibinfo{author}{\bibfnamefont{S.~K.} \bibnamefont{Reddy}},
  \bibinfo{author}{\bibfnamefont{N.~B.} \bibnamefont{Cramer}},
  \bibinfo{author}{\bibfnamefont{T.}~\bibnamefont{Cross}},
  \bibinfo{author}{\bibfnamefont{R.}~\bibnamefont{Raj}}, \bibnamefont{and}
  \bibinfo{author}{\bibfnamefont{C.~N.} \bibnamefont{Bowman}},
  \bibinfo{journal}{Chem. Mat.} \textbf{\bibinfo{volume}{15}},
  \bibinfo{pages}{4257} (\bibinfo{year}{2003}).

\bibitem[{\citenamefont{Wegscheider}(1923)}]{Wegscheider}
\bibinfo{author}{\bibfnamefont{R.}~\bibnamefont{Wegscheider}},
  \bibinfo{journal}{Z. Phys. Chem. CIII} \textbf{\bibinfo{volume}{103}},
  \bibinfo{pages}{273} (\bibinfo{year}{1923}).

\bibitem[{\citenamefont{Mauser}(1967)}]{Mauser}
\bibinfo{author}{\bibfnamefont{H.~Z.} \bibnamefont{Mauser}},
  \bibinfo{journal}{Naturforsch. B} \textbf{\bibinfo{volume}{22}},
  \bibinfo{pages}{569} (\bibinfo{year}{1967}).

\bibitem[{\citenamefont{Warren and Boettinger}(1995)}]{warren95}
\bibinfo{author}{\bibfnamefont{J.~A.} \bibnamefont{Warren}} \bibnamefont{and}
  \bibinfo{author}{\bibfnamefont{W.}~\bibnamefont{Boettinger}},
  \bibinfo{journal}{Acta Metall. Mater.} \textbf{\bibinfo{volume}{43}},
  \bibinfo{pages}{689} (\bibinfo{year}{1995}).

\bibitem[{\citenamefont{Ferreiro et~al.}(2002)\citenamefont{Ferreiro, Douglas,
  Warren, and Karim}}]{ferreiro02}
\bibinfo{author}{\bibfnamefont{V.}~\bibnamefont{Ferreiro}},
  \bibinfo{author}{\bibfnamefont{J.~F.} \bibnamefont{Douglas}},
  \bibinfo{author}{\bibfnamefont{J.~A.} \bibnamefont{Warren}},
  \bibnamefont{and} \bibinfo{author}{\bibfnamefont{A.}~\bibnamefont{Karim}},
  \bibinfo{journal}{Phys. Rev. E} \textbf{\bibinfo{volume}{65}},
  \bibinfo{pages}{051606} (\bibinfo{year}{2002}).

\bibitem[{\citenamefont{Lee and McKenna}(1988)}]{MCKENNA}
\bibinfo{author}{\bibfnamefont{A.}~\bibnamefont{Lee}} \bibnamefont{and}
  \bibinfo{author}{\bibfnamefont{G.~B.} \bibnamefont{McKenna}},
  \bibinfo{journal}{Polymer} \textbf{\bibinfo{volume}{29}},
  \bibinfo{pages}{1812} (\bibinfo{year}{1988}).

\bibitem[{\citenamefont{Rytov et~al.}(1966)\citenamefont{Rytov, Ivanov, Ivanov,
  and Anisimov}}]{rytov}
\bibinfo{author}{\bibfnamefont{B.~L.} \bibnamefont{Rytov}},
  \bibinfo{author}{\bibfnamefont{V.~B.} \bibnamefont{Ivanov}},
  \bibinfo{author}{\bibfnamefont{V.~V.} \bibnamefont{Ivanov}},
  \bibnamefont{and} \bibinfo{author}{\bibfnamefont{V.~M.}
  \bibnamefont{Anisimov}}, \bibinfo{journal}{Polymer}
  \textbf{\bibinfo{volume}{37}}, \bibinfo{pages}{5695} (\bibinfo{year}{1966}).

\bibitem[{\citenamefont{Terrones and
  Pearlstein}(2001{\natexlab{a}})}]{terrones1}
\bibinfo{author}{\bibfnamefont{G.}~\bibnamefont{Terrones}} \bibnamefont{and}
  \bibinfo{author}{\bibfnamefont{A.~J.} \bibnamefont{Pearlstein}},
  \bibinfo{journal}{Macromolecules} \textbf{\bibinfo{volume}{34}},
  \bibinfo{pages}{3195} (\bibinfo{year}{2001}{\natexlab{a}}).

\bibitem[{\citenamefont{Terrones and
  Pearlstein}(2001{\natexlab{b}})}]{terrones2}
\bibinfo{author}{\bibfnamefont{G.}~\bibnamefont{Terrones}} \bibnamefont{and}
  \bibinfo{author}{\bibfnamefont{A.~J.} \bibnamefont{Pearlstein}},
  \bibinfo{journal}{Macromolecules} \textbf{\bibinfo{volume}{34}},
  \bibinfo{pages}{8894} (\bibinfo{year}{2001}{\natexlab{b}}).

\bibitem[{\citenamefont{Terrones and Pearlstein}(2003)}]{terrones3}
\bibinfo{author}{\bibfnamefont{G.}~\bibnamefont{Terrones}} \bibnamefont{and}
  \bibinfo{author}{\bibfnamefont{A.~J.} \bibnamefont{Pearlstein}},
  \bibinfo{journal}{Macromolecules} \textbf{\bibinfo{volume}{36}},
  \bibinfo{pages}{6346} (\bibinfo{year}{2003}).

\bibitem[{\citenamefont{Terrones and Pearlstein}(2004)}]{terrones4}
\bibinfo{author}{\bibfnamefont{G.}~\bibnamefont{Terrones}} \bibnamefont{and}
  \bibinfo{author}{\bibfnamefont{A.~J.} \bibnamefont{Pearlstein}},
  \bibinfo{journal}{Macromolecules} \textbf{\bibinfo{volume}{37}},
  \bibinfo{pages}{1565} (\bibinfo{year}{2004}).

\bibitem[{\citenamefont{Ivanov and Decker}(2001)}]{Ivanov}
\bibinfo{author}{\bibfnamefont{V.~V.} \bibnamefont{Ivanov}} \bibnamefont{and}
  \bibinfo{author}{\bibfnamefont{C.}~\bibnamefont{Decker}},
  \bibinfo{journal}{Polymer Int.} \textbf{\bibinfo{volume}{50}},
  \bibinfo{pages}{113} (\bibinfo{year}{2001}).

\bibitem[{\citenamefont{Goodner and Bowman}(2002)}]{goodner02}
\bibinfo{author}{\bibfnamefont{M.~D.} \bibnamefont{Goodner}} \bibnamefont{and}
  \bibinfo{author}{\bibfnamefont{C.~N.} \bibnamefont{Bowman}},
  \bibinfo{journal}{Chem. Eng. Sci.} \textbf{\bibinfo{volume}{57}},
  \bibinfo{pages}{887} (\bibinfo{year}{2002}).

\bibitem[{\citenamefont{O'Brien and Bowman}(2003)}]{obrien}
\bibinfo{author}{\bibfnamefont{A.}~\bibnamefont{O'Brien}} \bibnamefont{and}
  \bibinfo{author}{\bibfnamefont{C.~N.} \bibnamefont{Bowman}},
  \bibinfo{journal}{Macromolecules} \textbf{\bibinfo{volume}{36}},
  \bibinfo{pages}{7777} (\bibinfo{year}{2003}).

\bibitem[{\citenamefont{Miller et~al.}(2002)\citenamefont{Miller, Gou,
  Narayanan, and Scranton}}]{Miller}
\bibinfo{author}{\bibfnamefont{G.~A.} \bibnamefont{Miller}},
  \bibinfo{author}{\bibfnamefont{L.}~\bibnamefont{Gou}},
  \bibinfo{author}{\bibfnamefont{V.}~\bibnamefont{Narayanan}},
  \bibnamefont{and} \bibinfo{author}{\bibfnamefont{A.~B.}
  \bibnamefont{Scranton}}, \bibinfo{journal}{J. Polym. Sci. A, Polym. Chem.
  Ed.} \textbf{\bibinfo{volume}{40}}, \bibinfo{pages}{793}
  (\bibinfo{year}{2002}).

\bibitem[{\citenamefont{Belk et~al.}(2003)\citenamefont{Belk, Kostarev,
  Volpert, and Yudina}}]{belk}
\bibinfo{author}{\bibfnamefont{M.}~\bibnamefont{Belk}},
  \bibinfo{author}{\bibfnamefont{K.~G.} \bibnamefont{Kostarev}},
  \bibinfo{author}{\bibfnamefont{V.}~\bibnamefont{Volpert}}, \bibnamefont{and}
  \bibinfo{author}{\bibfnamefont{T.~M.} \bibnamefont{Yudina}},
  \bibinfo{journal}{J. Phys. Chem. B} \textbf{\bibinfo{volume}{107}},
  \bibinfo{pages}{10292} (\bibinfo{year}{2003}).

\bibitem[{\citenamefont{Hirose et~al.}(1990)\citenamefont{Hirose, Wakasa, and
  Yamaki}}]{Hirose}
\bibinfo{author}{\bibfnamefont{T.}~\bibnamefont{Hirose}},
  \bibinfo{author}{\bibfnamefont{K.}~\bibnamefont{Wakasa}}, \bibnamefont{and}
  \bibinfo{author}{\bibfnamefont{M.}~\bibnamefont{Yamaki}},
  \bibinfo{journal}{J. Mater. Sci.} \textbf{\bibinfo{volume}{25}},
  \bibinfo{pages}{1209} (\bibinfo{year}{1990}).

\bibitem[{\citenamefont{Gumbel}(1958)}]{gumbel}
\bibinfo{author}{\bibfnamefont{E.~J.} \bibnamefont{Gumbel}},
  \emph{\bibinfo{title}{Statistics of Extremes}} (\bibinfo{publisher}{Columbia
  University Press, New York}, \bibinfo{year}{1958}).

\bibitem[{\citenamefont{Ben-Naim et~al.}(2001)\citenamefont{Ben-Naim,
  Krapivsky, and Majumdar}}]{majumdar01}
\bibinfo{author}{\bibfnamefont{E.}~\bibnamefont{Ben-Naim}},
  \bibinfo{author}{\bibfnamefont{P.~L.} \bibnamefont{Krapivsky}},
  \bibnamefont{and} \bibinfo{author}{\bibfnamefont{S.~N.}
  \bibnamefont{Majumdar}}, \bibinfo{journal}{Phys. Rev. E}
  \textbf{\bibinfo{volume}{64}}, \bibinfo{pages}{035101}
  (\bibinfo{year}{2001}).

\bibitem[{\citenamefont{Majumdar}(2003)}]{majumdar02}
\bibinfo{author}{\bibfnamefont{S.~N.} \bibnamefont{Majumdar}},
  \bibinfo{journal}{Phys. Rev. E} \textbf{\bibinfo{volume}{68}},
  \bibinfo{pages}{026103} (\bibinfo{year}{2003}).

\end{thebibliography}
\end{document}